\documentclass{aastex}
\usepackage{psfig}
\begin{document}
\newcommand{\kms}{km~s$^{-1}$}
\newcommand{\Msun}{M_{\odot}}
\newcommand{\ML}{M_{\odot}/L_{\odot}}
\newcommand{\etal}{{et al.}\ }
\newcommand{\hhh}{h_{100}}
\newcommand{\hsq}{h_{100}^{-2}}
\newcommand{\tn}{\tablenotemark}
\newcommand{\mdot}{\dot{M}}
\newcommand{\p}{^\prime}



\title{Associations of Dwarf Galaxies}

\author{R. Brent Tully}
\affil{Institute for Astronomy, University of Hawaii, Honolulu, HI 96822}

\author{L. Rizzi}
\affil{Institute for Astronomy, University of Hawaii, Honolulu, HI 96822}

\author{A. E. Dolphin}
\affil{Steward Observatory, University of Arizona, Tucson, AZ 85721}

\author{I.D. Karachentsev}
\affil{Special Astrophysical Observatory, Nizhnij Arkhyz, 369167, 
Karachaevo-Cherkessia, Russia}

\author{V.E. Karachentseva}
\affil{Astronomical Observatory, Kiev University, 04053 Kiev, Ukraine}

\author{D.I. Makarov\altaffilmark{1,2,3},L. Makarova\altaffilmark{1,2}}
\affil{Institute for Astronomy, University of Hawaii, Honolulu, HI 96822}

\author{S. Sakai}
\affil{Division of Astronomy and Astrophysics, University of California
at Los Angeles, Los Angeles, CA 90095-1562}

\and

\author{E.J. Shaya}
\affil{Astronomy Department, University of Maryland, College Park, MD 20743}

\altaffiltext{1}{also Special Astrophysical Observatory of the Russian
Academy of Sciences,  Nizhnij Arkhyz, 369167, Karachaevo-Cherkessia,
Russia}
\altaffiltext{2}{Isaac Newton Institute of Chile, SAO Branch}
\altaffiltext{3}{Observatoire de Lyon, 9, avenue Charles Andr\'e,
69561, St-Genis Laval Cedex, France}

\begin{abstract} 
Hubble Space Telescope Advanced Cameras for Surveys  has been used to 
determine accurate distances for 20 galaxies from measurements of the
luminosity of the brightest red giant branch stars.
Five associations of dwarf galaxies that had originally been identified 
based on strong correlations on the plane of the sky and in velocity are
shown to be equally well correlated in distance.  Two more associations
with similar properties have been discovered.  Another association is
identified that is suggested to be unbound through tidal disruption.
The associations have
the spatial and kinematic properties expected of bound structures 
with $1 - 10 \times 10^{11}~\Msun$.  However, these entities have little
light with the consequence that mass-to-light ratios are in the range 
$100 - 1000~\ML$.  Within a well surveyed volume extending to 3~Mpc, 
all but one known galaxy lies within one of the groups or associations
that have been identified.
\end{abstract} 

\keywords{galaxies: distances,
          galaxies: clusters,
          dark matter
}

\section{Introduction} 

The possibility was raised by \cite{tul02} [TSTV] that bound associations
of dwarf galaxies may be common.  In that paper, five interesting examples
were identified that were thought to be within $\sim 5$~Mpc.  
In terms of dimensions and velocity dispersions the entities
resemble scaled down versions of familiar nearby loose groups like the
Local Group.  They have dimensions of a few hundred kpc and velocity
dispersions of a few tens of \kms, implying group masses of 
$1-6 \times 10^{11} \Msun$.  However there is little light.  The implied
mass-to-light ratios ($M/L$) for four of the five entities identified by
TSTV were very large.

The identification of dwarf galaxy associations goes back two decades
to a project that defined groups through a merging tree algorithm
\citep{tul87, tul88}.  Each linkage between galaxies was characterized 
by a luminosity
density given by the separations and summed luminosities of the
contributing systems.  Two levels of structure -- `groups' and 
`associations' -- were defined by luminosity density thresholds.  The
`group' threshold satisfactorily captured familiar loose groups like
the Local Group.
The `association' threshold, set an order of magnitude lower in luminosity
density, captured two kinds of entities.  Associations of type 1 
involve the extended regions around groups defined by the higher luminosity
density threshold and can include several groups and/or several
individual galaxies (call these `associations of group peripheries').  
Associations of type 2 are derived from 
linkages between galaxies that have such insignificant luminosities that the
luminosity density fails to reach the threshold to be called a group (call
these `associations of dwarfs').  
It is this latter
kind of association that interests us in the present discussion.  Only 
limited attention was brought
to these entities when they were identified in 1987 because to suggest that
they were bound was to suggest that $M/L$ values are extreme.  TSTV asked
belatedly that we consider this possibility.

Except for one nearby case, the TSTV associations were identified in the
absence of distance information.  The candidate members simply lie near
each other in projection and have similar redshifts.  Thanks to the amazing 
capability of Advanced Cameras for Surveys (ACS) on Hubble Space Telescope
(HST) to detect faint stars, it has been possible to determine accurate
distances to all the suspected members of the TSTV dwarf associations.
We use the standard candle property of the tip of the red giant branch
(TRGB) \citep{lee93, sak96, mak06, riz06}.  In this paper, we revisit the 
discussion
of the putative dwarf groups, now with excellent distances in hand. 

\section{HST Observations}

The observations that will be reported here are part of a major program
to define the nearby structure in the distribution of galaxies.  The genesis
of the program arose out of all-sky searches for dwarf galaxies and
follow up HI observations that established redshifts \citep{fis81,kaa98,kaa99,
kar00,kaa00,huc01}.  In recent years, members of our team have reported
on observations of the resolved stellar populations in nearby galaxies made 
with HST \citep{kar02a,kar02b,kar02c,kar03a,kar03b,kar06,kar06b}, summarized 
in the Catalog of Neighboring Galaxies \citep{kar04}.  

Accurate distances can be
derived from the observed luminosity of the brightest red giant branch stars
in a galaxy.  Red giant stars increase in brightness while their Helium
cores grow until they attain sufficient mass that they cannot be supported 
by electron 
degeneracy pressure and begin burning to Carbon.  The well specified mass of 
the core
at this point results in a well defined TRGB luminosity \citep{ibe83}.
It has been empirically found that the tip luminosity is particularly 
stable for metal poor systems ([Fe/H] $< -0.7$ dex) at $I$ band, 
with $M_I \sim -4.05$ \citep{lee93}.  The location of the TRGB can be 
identified with 
sobel filter edge detection methods \citep{sak96} or maximum likelihood
methods \cite{men02}. Our preferred procedure is an extension of the maximum
likelihood method \citep{mak06}.  The absolute value of the TRGB and influences
that might cause it to vary have been discussed over the years
\citep{dac90,sal98,bar04}.  The calibration that we use and current 
uncertainties are discussed by \cite{riz06}.  For this paper we accept
$M_I = -4.05$ and no zero point offset for the HST flight filter F814W
magnitudes with either WFPC2 or ACS.

The bulk of the observations reported here were obtained with ACS during 
HST cycle 13, program 10210.  Some raw material was drawn from the HST 
archives as needed to complete the sample.  ACS data was reduced with the
DOLPHOT program \citep{dol06} while WFPC2 data was reduced with HSTPHOT
\citep{dol00}.  The TRGB fitting uses the modified maximum likelihood
method of \cite{mak06}.
A montage of images of the 20 galaxies observed in HST program 10210 is 
presented in Figure~\ref{fig:montage}.

Our current observations present the opportunity to show the spectacular 
advance over WFPC2 data made possible by ACS.  Figure \ref{fig:wfpc2-acs} 
compares
single HST orbit observations made by the two instruments on identical
galaxies.  The gain 
with ACS over WFPC2 is better than
$1^m$ with the same exposure time.  Another 0.4$^m$ gain comes from
a factor two longer exposure times.
In the case of UGC~8651, the color-magnitude diagram (CMD) obtained from the
ACS observations contains a well resolved red giant branch with the TRGB
clearly located.  One also sees the main sequence, asymptotic giant branch 
(AGB), and red supergiants.  These populations are also seen in the
WFPC2 CMD although not so cleanly.  A good distance measurement of 
$3.0 \pm 0.2$~Mpc can be made
for UGC~8651 with both datasets \citep{mak06}.  The same cannot confidently
be said in the case of KK~16.  The ACS observation is clean with a very
dominant RGB, and weak representation of the main sequence, red
supergiants, and AGB.  We measure a good distance of 
$5.5 \pm 0.3$~Mpc with
the ACS data.  If only the WFPC2 data were available the distance uncertainty 
would be large.  KK~16 is a relatively easy case because the young population
is a minor contributor to the CMD.  A much more difficult case is shown in
the bottom panels.  The ACS data tells us that UGC~3974 is at a distance of 
$8.0 \pm 0.8$~Mpc.  In the ACS CMD
we see very strong contamination from young stars.  Note the subtle difference
between the WFPC2 CMDs of KK~16 and UGC~3974, the latter with large main
sequence and supergiant populations which signal us to expect a substantial
presence of AGB stars.
We realize that the clump of faint red stars in the UGC 3974 WFPC2 CMD are 
all AGB stars and
that the TRGB is lost at the faint cutoff.  From the analysis of \cite{mak06}
it is concluded that single orbit observations with WFPC2 provide distances 
that can be trusted out to $\sim 5$~Mpc while ACS provides comparable
reliability out to $\sim 10$~Mpc.

Table~1 contains TRGB and distance determinations for all galaxies relevant
to the current discussion.  Entries are either (1) suspected members of one
of the associations that are discussed, and/or (2) lie in the well surveyed
volume with $\vert b \vert > 30$ and distance between 1.1 and 3.2 Mpc.
Column information includes association name, galaxy names (common and PGC),
equatorial, galactic, and supergalactic coordinates, blue apparent magnitude,
morphological type, velocity in the Local Group frame, HST program that 
provides the CMD, the TRGB magnitude at I band, foreground obscuration at
I band, distance modulus, distance (Mpc), velocity of galaxy minus mean 
velocity of association, 3-D distance of galaxy from centroid of association
(kpc), and absolute blue magnitude.

\section{Previously Identified Associations}

Prospective bound associations were identified by TSTV and \cite{tul05};
hereafter T05.
In the following discussion, the numeric names originate with the
Nearby Galaxies (NBG) catalog \citep{tul88} or are derivatives of that source.
The first number identifies the cloud or filament that contains the 
association -- usually 14, the one we live in.  The second number 
identifies the specific entity and is 
preceded by a negative sign for groups and a positive sign for associations.
As mentioned above, the difference between `groups' and `associations' in
the NBG catalog was a factor ten in luminosity density threshold.  In the 
present discussion we continue to use the designation `association' to
describe entities with very low luminosity densities even though we will
argue that these structures are bound, hence, dynamically, not very different
from entities like the Local Group.

\subsection{14+12 (NGC 3109) Association}

The NBG catalog used the 14+12 identification for all galaxies on the 
periphery of 14-12, our Local Group.  We now use the name for the dominant 
subset of these immediate neighbors; galaxies including NGC~3109 that lie
in the constellations of Sextans and Antlia.  These same 
objects were constituted as a separate group by \cite{vdb99}.  TSTV considered
6 candidates for membership but DDO~155 = GR8 is too far away in both
projection and distance to be retained and LSBC~D634-03 turned out to be
a distant dwarf projected onto a high velocity cloud concentration.
On the other hand, a new candidate for membership has recently turned up,
KKH~60.  This new candidate is near Sex~B in position and velocity 
\citep{mak03} but remains without a distance measure.

It was already known to TSTV that NGC~3109, Sex~A, and Sex~B are at
similar distances based on HST CMDs and that Antlia is at a similar distance 
based 
on groundbased CMD \citep{apa97,whi97}.  Now we have observed Antlia with
HST.
Figure~\ref{fig:14+12} shows the one new and three old HST CMDs.  The
TRGB are clearly defined and distances are confined to the remarkably 
small interval of $1.25 - 1.44$~Mpc. The rms scatter about the mean of 
1.37~Mpc is only 90~kpc.  The standard deviation in position on the sky
is much larger: 320~kpc.  The scatter in velocity (Local Group standard of 
rest) is only 18~\kms.  The rms line-of-sight velocity dispersion for this 
group 
is adjusted for observational uncertainties following \cite{mat74}, 
as it will be for all the groups to be discussed.
Individual observational uncertainties are taken to be $\pm 5$~\kms\ for 
these dwarf 
galaxies with narrow HI profiles so corrections to group dispersions are
only $\sim 1$~\kms.

The discussion of mass estimations will be put off to a later section.
Properties of the association are summarized in Table~2.  The brightest 
galaxy in this association is NGC~3109 with $M_B=-15.5$.
The 14+12 = NGC~3109
Association is the nearest distinct structure of multiple galaxies to the 
Local Group.

\subsection{14+8 Association}

For \cite{tul87}, the close trio UGC~8651, UGC~8760, and UGC~8833 was the 
archetype of associations of dwarfs.  UGC~9240 is near enough to these 
three that it must
be considered as a fourth member.  In TSTV, before distances were known,
it was guessed based on velocities that these galaxies would be at 
$\sim 5$~Mpc.  They turn out to be significantly closer, at 3.06~Mpc with a
scatter of only 210~kpc.  This compares with the scatter in projection of
220~kpc.  The scatter in velocity is 11~\kms.  The galaxies are all dwarfs
with $-12.4>M_B>-14.2$.  
CMDs are shown in Figure~\ref{fig:14+8}.

\subsection{17+6 (NGC~784) Association}

At the time of the construction of the NBG catalog, the 14 and 17 clouds were
seen as discrete but they are probably a continuation of the same filament.
The NBG catalog identifies only the pair NGC~784 and UGC~1281 with 
association 17+6.  KK~16 and KK~17 have subsequently been found.
NGC~784 and UGC~1281 are galaxies comparable to the SMC with $M_B=-16.6$
and -15.9 respectively and with prominent Pop I features in the CMDs.
KK~16 and KK~17 are dwarfs with only weak Pop I signatures.  
The distance of 5.2~Mpc is somewhat larger than anticipated by TSTV.
The scatter in distance is 230~kpc.  This compares with the projected scatter 
of 170~kpc.
The velocity dispersion is 17~\kms.

UGC~685 is located 1.3~Mpc from the centroid of the association and is not 
expected to be a bound member.
CMDs for members of the association and this outlier are shown in 
Figure~\ref{fig:17+6}.

\subsection{14+19 Association}

In this case, the NBG catalog identification with the 14 cloud is probably
not warranted.  The connection with the local 14 cloud was based on the
small mean velocity of 182~\kms.
The large mean distance of 7.9~Mpc suggests
these galaxies are probably part of the 15 cloud (the Leo Spur).
The low velocity and large distance
implies a peculiar velocity in the line-of-sight of --350 to --400~\kms.  
The Leo Spur generally manifests large motions toward us
(the `local velocity anomaly' \citep{tul92}).  Only two of the four 
candidate galaxies in the 14+19 Association
had been identified when the NBG catalog was published.

These galaxies lie beyond the range of WFPC2 SNAP observations and 
published distance estimates based
on WFPC2 observations were picking up the onset of the AGB rather than the 
RGB.  With ACS we adequately distinguish these separate features
(see lower panels of Fig.~\ref{fig:wfpc2-acs}).
Even with the increased uncertainty in the TRGB measurement so near the
photometric limit, the observed
scatter in the distances of the four galaxies is only 290~kpc, 4\% of the
mean distance.  The scatter in projected separations is 560~kpc.
Velocity dispersion is 22~\kms.  The brightest galaxy is UGC~3974 with
$M_B=-16.0$.  The association lies at low Galactic latitudes but there is
fortuitously low obscuration in the region.
 
The large distance found for the 14+19 association puts the membership of
UGC~3755 in doubt.  This galaxy is 1~Mpc from the centroid of the group.
The issue of its membership will be reconsidered in Section 5.

It is to be noted that an interim discussion of the dynamics of this
association by T05 was based on distances obtained from WFPC2.
As mentioned, those distance measures are badly underestimated.
CMDs are shown in Figure~\ref{fig:14+19}.

\subsection{14+13 (NGC~55) Association}

When the NBG catalog came out the principal members of this entity, NGC~55
and NGC~300, were taken to be part of the Sculptor Group, 14-13 in the NBG
catalog, with NGC~253 as the dominant galaxy.  It subsequently became
evident that NGC~55 and NGC~300 are to the foreground.  Three other galaxies
are associated: ESO~294-10, ESO~407-18, and ESO~410-05.  The latter, at
a distance of 1.94~Mpc, is a dE system with a velocity that was measured
only recently \citep{bou05}.  The mean
distance for the five galaxies is 2.07~Mpc, with a dispersion of 125~kpc.
The dispersion in projection is 230~kpc.  
Another galaxy, IC~5152,
has essentially the same distance (1.97~Mpc) and velocity ($V_{LG}=75$~\kms) 
but is
830~kpc away in projection.  The status of this outlier will be considered during the
discussion of the association dynamics.

The newly determined velocity of $V_{LG} = 176$~\kms\ for ESO~410-05 causes
a dramatic revision of our understanding of the dynamics of the 14+13
Association.  The line-of-sight rms velocity dispersion jumps from
$\pm 15$~\kms\ with 4 systems to $\pm 36$~\kms\ with 5 systems, with obvious
implications for the mass.

The velocity dispersion in this case, as with all the other associations 
discussed in this paper, is in the Local Group standard of rest \citep{kar96}.
Members of this nearby association subtend $\sim 20^{\circ}$ on the sky
and it can be asked if the velocity dispersion can be reduced by a correction
for a proper motion of the association centroid.  The possibility was pursued 
in this case because it could be seen that there was a gradient in $V_{LG}$ 
of $\sim 37$~km~s$^{-1}$~degree$^{-1}$ along an axis with an orientation
that could be defined to $\sim \pm 25^{\circ}$.
However the amplitude of motion
required to reduce the velocity dispersion of the association candidates
is ludicrous. 
A large transverse velocity of $2,000$~\kms\ would only
reduce the internal velocity dispersion of the association from
$38$~\kms\ to $29$~\kms.  The lesson learned was that any reasonable
proper motion has only a modest impact on the observed velocity dispersion.
The issue is less important for more distant associations because they
subtend smaller angles on the sky.

The 14+13 Association has the dimension and velocity dispersion 
characteristics of the entities discussed previously but it differs because
two of the galaxies are not dwarfs: NGC~55 with $M_B=-17.9$ and NGC~300
with $M_B=-17.7$.  As a consequence, it will be seen that this association 
has a lower mass to light ratio than the others and can be viewed as a 
transition structure
between associations of dwarfs and classical loose groups like the Local 
Group.
CMDs are shown for the candidate association members in Figure~\ref{fig:14+13}.

\section{Are There Any Other Nearby Dwarf Associations?}

The associations that have been discussed were considered by TSTV to be the
most interesting cases within a group averaged velocity of 400~\kms\
(ie, suspected to lie within $\sim 5$~Mpc; although we now find that the 
14+19 association is
substantially more distant at 8~Mpc and moving with a large peculiar motion
toward us!).  We took the opportunity with the HST observations to check
out other possibilities.  

\subsection{14+14 Association}

Our closest attention was given to the region 
around NGC~1313.  In the NBG catalog, NGC~1313 and ESO~115-21 constitute 
the 14-14 pair while a 14+14 association lies nearby to one side including
NGC~1311, IC~1959, and ESO~154-23.  The only new galaxy to come to light
that might belong with 14-14 is KK~27.  The CMDs for a possible
14-14 trio are shown in the upper part of Figure~\ref{fig:14+14} and for 
the 14+14 trio in the lower part of the same figure.

The distance information from the CMDd show that the 14+14 trio are distinctly
behind the 14-14 trio.  We infer that the two trios are dynamically distinct.
In the case of 14-14, 
KK~27 is found to be a dE companion to NGC~1313.  We do not have
a velocity for this galaxy so the kinematic information for the 14-14 
structure remains too sparse to be given attention.

The 14+14 trio have the joint properties that resemble the high mass end of 
our associations.
The velocity dispersion is relatively high at 35~\kms.  At a distance of
5.8~Mpc, the three galaxies
are all in the SMC luminosity range with $-15.5 > M_B > -16.2$.  The projected
inertial radius of 250~kpc compares with the rms difference in distance of
310~kpc.  Properties of the proposed association are summarized in Table~2.

\subsection{14+7 (NGC 4214) Association}

In the first systematic search for dwarf galaxies across the entire northern
sky, \cite{vdb66} found the largest concentration of prominent dwarfs was
in the constellation of Canes Venatici.  The nearer of these lie in a 
structure called Canes Venatici I by \cite{dev75} and the 14-7 Group in the
NBG catalog.  Now that excellent TRGB distances are becoming available,
it is becoming evident that this structure breaks up into distinct foreground
and background components that are completely confused in
velocity and position on the sky \citep{kar03b}.  The more luminous
galaxies, including NGC~4736, lie in the background component at 4.4~Mpc
and constitute the main 14-7 Group.  The foreground objects congegate 
around NGC~4214 at 2.8~Mpc and will henceforth be referred to as the 
14+7 Association.

At least six low luminosity systems lie in the foreground association, making 
it the best populated association of dwarfs known (these include NGC 3741,
4163, 4214 and DDO 99, 113, 125).  CMDs based on HST archival material are
presented for these galaxies in Figure~\ref{fig:14+7}.  The CMD for the 
outlyier UGC~8508 is shown in Figure~\ref{fig:dregs}.
Presently, only about half the galaxies in 
the Canes Venatici region with appropriate velocities have good distance 
measures so there is the prospect that the list of members of the 14+7
association will grow.  Accordingly, this association merits special
attention because it can provide better statistical information than the
other associations have afforded.  If additional members are identified
then they will improve our kinematic knowledge, hence mass estimate, but
they will not add much light.  The remaining candidates are all much less
luminous than NGC~4214 at $M_B = -17.1$.

\subsection{The dregs within 3 Mpc}

Most known galaxies within 3~Mpc and at high Galactic latitude now have
excellent distance estimates through the TRGB method.  Are there any 
interesting structures in the objects that are {\it not} assigned to 
previously identified condensations?  We consider all galaxies that are
beyond the traditional Local Group (taken to include Tucana and SagDIG at the
periphery) but within 3~Mpc in the compilation by \cite{kar04} or as updated by a few new
objects and improved TRGB detection algorithms \citep{mak06,kar06}.
There are 43 such galaxies.  Because of incompleteness problems we will 
not give attention to
the zone within 30 degrees of the Galactic plane except to say that there are 
13 known objects in or peripheral to the Maffei Group where the supercluster 
plane crosses the plane of our Galaxy in the north and 5 known objects in the 
region of Circinus and
the Centaurus Group at the southern crossing.  Probably other galaxies 
await discovery in this zone.  

Ignoring this obscured half of the sky
leaves 25 galaxies.  
One galaxy (KK~127) injected into our sample
strictly on the basis of its low velocity is probably a high negative
velocity member of the Coma~I Cluster at 16~Mpc. (Note: high negative 
velocity members of a similar nature in the Virgo Cluster had already been
discounted from consideration in the \cite{kar04} catalog.)
Then 19 galaxies belong to the 14+13, 14+12, 14+7, and front of the 14+8
dwarf associations that were discussed earlier, including the recently
discovered KKH~60 that does not have a distance measure.  Passing over these 
20
high latitude objects connected with already established structures takes the
count to be considered to only five, all with good distances.

One of these five, KKR~25, is well separated from the others.
It is nearby, with a TRGB measurement that puts it at a distance of 2.14~Mpc.
The CMD is seen in Fig.~\ref{fig:dregs}
A claimed velocity of $V_{helio} = -139$~km~s$^{-1}$ has not been confirmed
\citep{beg05} and is probably attributable to confusion with local HI.
Nonetheless, the distance to KKR~25 is reliable and
citizens of that dwarf spheroidal galaxy must see
the Milky Way as the nearest giant system.  It is a factor two more distant 
than the zero velocity surface separating Local Group infall from cosmic
expansion \citep{kar02a} but still within the zero energy surface
\citep{pei05} so conceivably is bound to the Local Group.   
Beyond it is the Local Void
\citep{tul87b}.  It is the most isolated galaxy known in this small volume
(it's nearest known neighbor is the dwarf KK~230 at a distance of 1.1~Mpc) 
and, remarkably, it is gas deficient!

What is to be made of the last four?  By name they are: 
UGC~8091 = DDO~155 = GR8, 
UGC~9128 = DDO~187, KKH~86, and KK~230 at a mean distance of 2.3~Mpc.  
Their CMDs are seen in Figure~\ref{fig:dregs}.
They are close to each other but not so close that we would expect them 
to be bound,
with an ensemble inertial radius in 3-dimensions of 570~kpc (inertial radius
$R_I = [\Sigma_i^N r_i^2 /N]^{1/2}$ where $r_i$ is the distance of galaxy $i$ 
from the
centroid of the N=4 objects).   The velocity dispersion of the four is 
37~\kms, at the upper limit of the associations of dwarfs previously 
discussed.  Turning these dimensions and velocities into
masses assuming a bound system implies values like 
$1 - 2 \times 10^{12}~\Msun$, comparable to the Local Group.  However the 
observed galaxies are all extreme dwarfs in the range $-8>M_B>-13$.  
If bound, it would require that $M/L_B \sim 50,000~\Msun$.

We do not expect that these four galaxies are bound together.  There is a 
monotonic increase in the velocity of each galaxy with distance from us, 
the property expected if the objects are in relative expansion.
These 4 galaxies might be considered as an `unbound association'. 
That 4 of 5 `leftovers' in a volume of $\sim 50 {\rm Mpc}^3$ at high 
latitudes within 3~Mpc should be contained within a volume of $1 {\rm Mpc}^3$
suggests some sort of linkage.  We will revisit the matter in the next section 
with the suggestion that this region is being tidally disrupted.

\section{Masses of the Associations Revisited}

Let it be understood clearly that we do not expect that the associations
of dwarfs are in dynamical equilibrium.   Crossing times can be 80\% of 
the Hubble
time, H$_0^{-1}$.  Guidance regarding the dynamical state of the 
structures can come from consideration of the nearest
generally accepted groups like the Local Group including M31 and the
Milky Way, the M81 Group including M81 and the NGC~2403 sub-region 
\citep{kar02a},
and the Centaurus Group including Cen~A and M83 as distinct dynamical 
centers \citep{kar06b}.  In each of these well known cases, there are 
dynamically evolved regions around 
major galaxies, marked by a predominance of early type companions and
motions that are both positive and negative with respect to the dominant
host.  In each case, too, there are more extended, lower density, relatively 
unevolved
regions, marked by predominantly late morphological types and infall velocity 
patterns.
The associations of dwarfs resemble the latter, low density regions, 
lacking a semblance of a core.

The zero-velocity surface or radius of first turnaround for our Local Group
occurs at about 900~kpc from the centroid of the
Milky Way and Andromeda galaxies \citep{kar02c,tul05b}.
One can scale to the
equivalent surface for a dwarf association mass 
by using the property that turnaround today occurs at the same density
everywhere.
Hence in the spherical approximation, $r_{dw} = r_{LG} (M_{dw}/M_{LG})^{1/3}$.
With $M_{LG}=2 \times 10^{12}~\Msun$ and $r_{LG}^{1t}=900$~kpc, 
then 
\begin{equation}
r_{dw}^{1t} \sim 330 M_{11}^{1/3}~{\rm kpc}
\end{equation}
where $M_{11}$ is the mass of the association halo in units of 
$10^{11}~\Msun$.  This scaling relation provides a rough requirement 
to be met as we evaluate whether an association might be bound or whether 
an individual galaxy is a serious candidate for membership.

Results for the ensemble of associations are summarized in Table~2.  
As a general comment, it will 
be noted that most of the $M/L$ values are lower than reported by TSTV
and T05.  Mostly, these reductions result from
better distances.  In addition, the earlier work had considered velocities 
in the 
Galactic standard of rest \citep{dev91} while here velocities are considered
in the Local Group standard of rest \citep{kar96}, removing gradients
caused by the motion of our Galaxy toward M31.  Most velocity
dispersions are reduced a bit as a consequence, hence inferred masses are 
reduced. 

Following T05, masses are calculated in two ways, the first 
making use of the ``projected mass estimator'' of \cite{hei85} and the 
second based on the virial theorem without luminosity weighting.  There 
was a lengthy discussion in T05 regarding mass estimate uncertainties
which are considerable.  The factor of 2 difference between the virial
mass estimate and the estimate that arises from the simple assumption
that the structure is bound gives a rough measure of systematic 
uncertainties.  Probably a better measure of uncertainties is given
by tracking how mass and $M/L$ values for a given entity have bounced around
between TSTV, T05, and the present results presented in Table~2.  The
median variation for a given entity is a factor 3, simply arising from
incremental improvements in distances, velocities, luminosities, and
membership assignments. 

Figure~\ref{fig:dv6} shows relative velocities as a function of relative 
distances
for the galaxies identified with the seven associations.  If gravity played
no role then one would expect a Hubble law dependence between velocity
and distance.  If motions were predominantly radial infall then one
could expect nearer galaxies in a group to be systematically redshifted 
compared with farther galaxies in the group.  There is a slight hint of a
positive correlation between velocity and distance in Fig.~\ref{fig:dv6}
but it comes exclusively from the 14+14 and 17+6 associations.  
The uncertainty in the distribution of points in this figure is too great
to warrant a dynamical interpretation.
It is to be
appreciated that overall expansion or contraction are subtle effects 
given the very small dispersions in velocities and significant errors in
distances.
Here are comments on each of the proposed associations in turn.  

\noindent
{\it 14+12 (NGC 3109) Association.}  
In TSTV and T05, the velocity dispersion was measured in the Galactic
rest frame.  With the transform from
Galactic to Local Group rest frame the velocity dispersion increases 
slightly, resulting in larger mass estimates than 
previously.
The new candidate KKH~60 does not yet have a distance measure.  The 
mass calculations include this fifth candidate placed at the mean group
distance and at $\pm 3 \sigma_d$ excursions from the mean distance where
$\sigma_d$ is the dispersion in the distances of the four established
members.  
The addition of the fifth candidate
causes mass estimates to drop 20\%, an insignificant amount
compared with other sources of error.  
Modifying the distance over the $\pm 3 \sigma_d$ range 
increase mass estimates by only $\sim 15\%$, a negligible amount.

The 14+12 Association is nearby the Local Group (1.37~Mpc from the Milky Way
and 1.7~Mpc from the Local Group mass centroid).  It can be asked if the 
association is stable against tidal disruption.  An
association mass of $\sim 1.6 \times 10^{11} \Msun$ would be stable against 
disruption from the Local Group with $\sim 1.8 \times 10^{12} \Msun$ out to
a sphere radius of $\sim 500$~kpc.  This dimension is larger than the dimension
of the association, so tidal disruption is not inferred to be a problem today,
although it is a close call.
It turns out that the 14+12 Association presents
the most vulnerable case from the standpoint of tidal disruption among the
associations that have been currently identified.  The possibility of tidal
disruption does have relevance in a few situations regarding outliers.

\noindent
{\it 14+8 Association.}
Mass and $M/L$ estimates dropped from T05 because of a lower velocity 
dispersion
in the Local Group frame and a 30\% increase in the attributed luminosity.
These changes provide a warning that the results have large uncertainties.
Still, $M/L$ values are an order of magnitude larger than values attributed
to nearby spiral rich groups \citep{kar05b}.

\noindent
{\it 17+6 (NGC 784) Association.}
The mass estimates dropped substantially from T05 for two reasons.  In this
case the velocity dispersion dropped dramatically passing to the Local Group
rest frame.  Moreover, with improved distances, the 3-dimensional inertial
radius is shown to be only 2/3 its previously assumed value.  

The dwarf galaxy UGC~685 lies 1.3~Mpc from the centroid of the 17+6
Association.  The zero-velocity surface for this association is inferred
from Eq.(1) to lie at $\sim 380$~kpc so UGC~685 must be not be bound with 
the others.

\noindent
{\it 14+19 Association.}
$M/L$ estimates are reduced, partially because of a reduced velocity 
dispersion and partially because of a significant increase in distance.
The virial mass estimator for this group is subject to significant error 
because of the proximity of
UGC~3974 and KK~65.  The virial mass estimator is unstable if numbers are
small and there are instances of close pairs.

At a distance of almost 8~Mpc, we are forced to reevaluate if UGC~3755 should 
be considered a member.  At this large distance, we find it is 1~Mpc from 
the centroid of the 4 candidates.
It is seen in Table~2 that the mass found for the 14+19 association is
$\sim 4 \times 10^{11}~\Msun$ whether 3 or 4 members are considered.
The zero velocity turnaround surface of a bound system would be at a
radius of $\sim 520$~kpc according to Eq.~(1). 
We conclude that UGC~3755 would not be within the association infall region.
Nonetheless, if we consider the model of spherical infall with
$\Omega_{\Lambda} = 0.7$ and $\Omega_m = 0.3$ discussed by \cite{pei05}, 
we infer that the bound surface of a mass of $4 \times 10^{11}~\Msun$
would lie at a radius of $\sim 1.2$~Mpc.  UGC~3755 could yet be trapped by 
the 14+19 Association.

\noindent
{\it 14+13 Association.}
This interesting case was discussed by TSTV and T05: an association 
with the same properties as the others in terms of dimensions and velocity 
dispersion but {\it not} deficient in light.  The addition of ESO 410-05
changes the picture.  This small galaxy is clearly a member based on its
coordinates and distance but its inclusion causes the velocity dispersion 
to jump by a factor 2.4 and mass estimates to jump by a factor 6.
Whereas previously this association stood apart with a low value of $M/L$, 
now with an increased mass estimate it
lies in an interesting transition regime between associations of
dwarfs and entities like the Local Group.

The 14+13 Association has an outlier:
IC~5152 is 820~kpc from
the centroid of the association, beyond the zero-velocity turnaround radius 
of $\sim 570$~kpc anticipated by Eq.~(1).
The tidal radius for the
14+13 Association ($1.6 \times 10^{11} \Msun$ at 2~Mpc from the Local Group
with $1.8 \times 10^{12} \Msun$) is at a radius of $\sim 1$~Mpc so IC~5152
is near the tidal surface.

\noindent
{\it 14+14 Association.}
This entity is a new addition.  With only 3 candidate members the derived
parameters are uncertain.  However they are within the range of the other
cases.

\noindent
{\it 14+7 Association}
Discussion of this entity is preliminary because other potential members
are identified but, lacking good distance measures, are entangled with 
the background 14-7 Group.  The six confirmed members already constitute
the largest association of dwarfs known.  The luminosity of the association
is reasonably established because of the dominance of NGC~4214 so the
high value of $M/L$ reported in Table~2 already has a firm basis.  A seventh
galaxy with a good distance, UGC~8508, lies $\sim 900$~kpc from the 
centroid of the association, undoubtedly beyond the infall region but
possibly bound (ie, within the zero-energy surface).

\noindent{\it The dregs.}
If we ignore intuition and consider the four galaxies UGC~8091 = DDO~155, 
UGC~9128 = DDO~187, KKH~86, and KK~230 to be bound then one gets the
results reported in Table~2.  The entity would bear closer resemblance to 
the Local Group
in terms of dimensions, velocity dispersion and inferred mass than to the
associations that have been discussed.  The combined
luminosity is much lower than any of the other cases, so the inferred
$M/L$ is much larger.  

As was noted, the distances of the four galaxies among the dregs increase
monotonically with velocity, the condition expected if the objects are
unbound.  Given the proximity of the Local Group, these galaxies would need 
at least $1.4 \times 10^{11} \Msun$ (whence $M/L > 5000 \ML$) to stabilize
against tidal disruption.
The most reasonable assumption is that the four galaxies are 
{\it not} bound and the dynamical analysis provides no information regarding
the distribution of mass in their vicinity.
We may be seeing a tidally disrupted, evaporating association.  Little 
can be said 
with a case of one.  Presently our three-dimensional census of dwarfs
and their distances
is very incomplete beyond 3~Mpc even at high Galactic latitudes.

\section{Discussion}

It is sobering to see the changes that have occurred in the inferred
masses and $M/L$ values for the putative associations of dwarfs since
they were identified by TSTV and rediscussed by T05.  The changes result
from updates in distances, velocities, luminosities, and membership
assignments.  Yet the fundamental conclusion remains the same.  
{\it If the associations are bound then $M/L$ values are very high.}

There was an extensive discussion in T05 in favor of the proposition that
the associations of dwarfs are bound and, if so, of the uncertainties 
that arise in mass estimates.  Here, we will briefly recall the two
arguments regarding the more important of these issues; that the associations
are bound.

The first argument notes the continuity of inferred dark matter halo
properties of the associations of dwarfs with more familiar groups of
spirals.  Figure~\ref{fig:rv}, a modification of a plot in T05, compares group 
velocity dispersions and projected inertial radii for both dwarf associations
and spiral groups.
A continuum of these properties are seen from spiral group velocity 
dispersions of $\sim 100$~\kms\ and radii of $\sim 600$~kpc down to dwarf 
associations at $\sim 25$~\kms\ and radii of $\sim 250$~kpc.
However, while there is a continuity in dynamical properties, whence
inferred mass properties, there is a break in the relationship with light
between the spiral and dwarf regimes.

The second argument develops out of the observation that dwarf galaxies are
highly correlated in position.  Figure~6 in T05 shows the 3--dimensional
2--point correlation for galaxies at distances between one and five Mpc 
at high Galactic latitude, demonstrating a strong peak on scales less than
500~kpc.  We will make the same point by looking at the latest data available 
for the restricted shell between 1.1 and 3.2 Mpc and $\vert b \vert >30$ where
there is a high level of completion.  The lower distance limit excludes
galaxies considered to be part of the Local Group.  The upper limit excludes
galaxies considered to be members of the nearest well known groups around
M81 and in Sculptor, Centaurus, and Canes Venatici.  There are 24 galaxies
with well established distances within this volume of 66~Mpc$^3$.  A small
number of additional galaxies identified but without measured distances are
plausibly within this volume. 

The clumping of these 24 galaxies is extreme.  Two of these, along with 
NGC~3109 and Antlia which are slightly below $b = 30$,
are at 1.4~Mpc in an enclosing spherical volume of 0.4~Mpc$^3$ 
(plus suspected companion KKH60).  Five, with NGC~55 the brightest, are at
2.1~Mpc in 0.3~Mpc$^3$ (plus sixth outlier IC~5152).  Six, with NGC~4214 the
brightest, are at 2.8~Mpc in 0.6~Mpc$^3$ (plus seventh outlier UGC~8508).
Four, with UGC~9240 the brightest, are at 3.1~Mpc in 0.4~Mpc$^3$.  Five more,
with UGC~9128 the brightest, at 2.3~Mpc are more loosely linked within 
1.4~Mpc$^3$.  Hence 23 of 24 galaxies with good distances lie in five regions
enclosing $\sim 3$~Mpc$^3$ of the available 66~Mpc$^3$.  The distribution of
these galaxies is shown first in Figure~\ref{fig:shell} in projection on
the sky and then in Figure~\ref{fig:xyz} in two projections in supergalactic
cartesian coordinates.  These galaxies inhabit only a small part of the
available volume.

We have presented two arguments that build a strong circumstantial case 
that dwarf galaxies in
associations, like galaxies in spiral-dominated groups, are born in overdense
regions that are bound.  In cases like the `dregs' including UGC~9128, and the
association outliers like IC~5152 and UGC~8508, it
is suspected that tidal forces have had a disrupting influence.
However, in the remaining cases we infer that the associations remain bound
and, consequently our effort to estimate masses has validity.

The results are illustrated in Figures \ref{fig:ml} and \ref{fig:mlm}.  
The data
in these figures is extracted from T05 except for updated values drawn
from the current paper for the associations of dwarfs and revised values
for the Local, M81, and Centaurus groups \citep{kar05b,kar06b}.  
The situation is
not significantly changed.  If associations of dwarfs are representative
of bound structures on mass scales below $10^{12}~\Msun$ then there is a
sharp break below this scale in the relationship between light and mass.
The expression fit to the data in Fig.~\ref{fig:ml} (minimizing the square
of deviations in mass) is:
\begin{equation}
L_B = \phi M^{\gamma} e^{M^{\dagger}/M}
\end{equation}
where the power law slope is $\gamma = 0.59$, the exponential cutoff is
characterized by $M^{\dagger} = 6 \times 10^{11}~\Msun$, and the 
normalization is $\phi = 2700$.

A similar minimum in $M/L$ with a sharp increase at lower masses has been
suggested by others.  The shallow faint end slope of observed luminosity 
functions of galaxies compared with the halo mass function has indicated
a paucity of light at low masses \citep{mar02,vdb03,vdb05}.  Semi-analytic
models have found a similar effect, arising out of feedback heating that 
damps the accumulation of gas and star formation in low mass halos
\citep{ben00}.  Those hints of a rapid increase of $M/L$ toward lower
masses are based on theoretical considerations.  The claims made in T05
and this paper are directly based on observations.

How has the present study clarified the situation?  It has convincingly
been shown that the strong correlations in projection and in velocity of
a large fraction of nearby dwarf galaxies are, in fact, strong
correlations in three-dimensions.  The associations of dwarfs have the
spatial distribution and kinematic properties expected of structures like
the Local Group but less massive by factors 3 to 10.  We infer that
these associations are bound but dynamically unevolved.  To make up the
implied masses, they presumably
contain dark matter sub-halos on scales of $10^9 - 10^{10}~\Msun$ that
contain too little gas and too few stars to be currently detected.

There is a strong incentive to increase the size of the census of nearby
galaxies.  We probably have identified all galaxies brighter than
$M_B \sim -10$ at $\vert b \vert > 30^{\circ}$ and within 3~Mpc.  Aside
from lonely KKR~25, every object in this volume is associated either with
a luminous group or an association of dwarfs or the `dregs' evaporating
association.  Our current knowledge is predominantly the consequence of a 
series of observations with HST WFPC2.  The added sensitivity of ACS
doubles the range available for accurate distance measurements.  It is within 
our
capability to get distances good to 5\% for all galaxies at high latitudes
with $M_B < -10$ and within 6--8~Mpc.  
With completion in a
volume an order of magnitude greater than currently available, the
statistics will improve and the case should become clear regarding
whether or not associations with little light are bound.

\vskip 1cm
\noindent
This research has been sponsored by STScI award HST-GO-10210.
DIM received support from INTAS grant 03--55--1754 and from the Russian 
Science Support Foundation.

\clearpage
\pagestyle{empty}
\begin{deluxetable}{llcrrrrrrrrrccrcr}
\rotate
\tabletypesize{\scriptsize}
\tablecaption{Candidate Association Members or Galaxies with $1.1 < d < 3.2$ Mpc
at $\vert b \vert >30$. \label{tbl-1}}
\tablewidth{0pt}
\tablehead{
\colhead{Group} & \colhead{Names} & \colhead{RA (J2000) Dec} & \colhead{SGL} &
\colhead{SGB} & \colhead{$B_T$}  & \colhead{Ty} & \colhead{$V_{LG}$} &
\colhead{HST} & \colhead{$I_{tip}$}  & \colhead{$\pm$} & \colhead{$A_I$} &
\colhead{$m-M$} & \colhead{$d$ (Mpc)} & \colhead{$\Delta V$} & \colhead{$R$ (kpc)} &
\colhead{$M_B$}
}
\startdata
14+12 &  NGC 3109, DDO 236   &  100306.6 -260932 &  262.0981 &  -45.1065 &  10.10 &   9 &   110 &   8601 &  21.71 &  0.03 &  0.13 &  25.63 &  1.34 &   15 &   307 &  -15.53 \\
      &  U5373, SexB, D 70   &  095959.8 +051957 &  233.1991 &  -39.6203 &  11.75 &  10 &   111 &   8601 &  21.80 &  0.02 &  0.06 &  25.79 &  1.44 &   16 &   450 &  -14.03 \\
      &  Sex A, DDO 75       &  101101.3 -044248 &  246.1704 &  -40.6659 &  11.94 &  10 &    94 &   7496 &  21.82 &  0.06 &  0.09 &  25.78 &  1.43 &   -1 &   212 &  -13.84 \\
      &  Antlia              &  100403.9 -272001 &  263.0984 &  -44.8031 &  16.19 &  10 &    66 &  10210 &  21.60 &  0.13 &  0.15 &  25.49 &  1.25 &  -29 &   346 &   -9.30 \\
      &                      &                   &           &           &        &     &       &        &        &       &       &        &       &      &       &         \\
14+13 &  NGC 55              &  001508.4 -391313 &  332.6740 &   -2.4102 &   8.78 &   9 &   111 &   8697 &  22.66 &  0.03 &  0.03 &  26.68 &  2.17 &   -5 &   186 &  -17.90 \\
      &  NGC 300             &  005452.6 -374057 &  299.2306 &   -9.4984 &   8.89 &   7 &   114 &   9492 &  22.52 &  0.03 &  0.03 &  26.54 &  2.04 &   -2 &   322 &  -17.65 \\
      &  UGCA 438, E407-18   &  232622.2 -322329 &   11.8694 &    9.2990 &  13.80 &  10 &    99 &   8192 &  22.72 &  0.06 &  0.03 &  26.74 &  2.22 &  -17 &   404 &  -12.94 \\
      &  ESO 410-005         &  001531.5 -321047 &  357.8469 &   -0.2573 &  14.85 &  -5 &   176 &   8192 &  22.42 &  0.05 &  0.03 &  26.44 &  1.94 &   60 &   196 &  -11.59 \\
      &  ESO 294-010         &  002633.3 -415119 &  320.4159 &   -5.2699 &  15.51 &  10 &    81 &   8601 &  22.42 &  0.08 &  0.01 &  26.46 &  1.96 &  -35 &   195 &  -10.95 \\
      &                      &                   &           &           &        &     &       &        &        &       &       &        &       &      &       &         \\
      &  IC 5152             &  220241.2 -511747 &  343.9191 &   11.5285 &  10.95 &  10 &    75 &   8192 &  22.47 &  0.08 &  0.05 &  26.47 &  1.97 &  -41 &   817 &  -15.52 \\
      &                      &                   &           &           &        &     &       &        &        &       &       &        &       &      &       &         \\
14+07 &  NGC 4214            &  121538.6  361941 &  160.2577 &    1.5936 &  10.15 &  10 &   295 &   6569 &  23.32 &  0.03 &  0.04 &  27.33 &  2.92 &   46 &   179 &  -17.18 \\
      &  UGC 7577, DDO 125   &  122740.9  432944 &  137.7581 &    5.9272 &  12.86 &  10 &   240 &   8601 &  23.18 &  0.03 &  0.04 &  27.19 &  2.74 &   -9 &   381 &  -14.33 \\
      &  NGC 4163            &  121209.1  361009 &  163.2040 &    0.8761 &  13.54 &  10 &   164 &   9771 &  23.33 &  0.02 &  0.04 &  27.34 &  2.94 &  -85 &   227 &  -13.81 \\
      &  UGC 6817, DDO 99    &  115053.0  385249 &  166.1978 &   -2.1220 &  13.59 &  10 &   248 &   8601 &  23.11 &  0.04 &  0.05 &  27.11 &  2.64 &   -1 &   294 &  -13.52 \\
      &  NGC 3741            &  113606.2  451701 &  157.5712 &   -2.0769 &  14.28 &  10 &   264 &   8601 &  23.52 &  0.06 &  0.05 &  27.52 &  3.19 &   15 &   510 &  -13.24 \\
      &  UGCA 276, DDO 113   &  121457.9  361308 &  161.1018 &    1.4295 &  15.61 &  10 &   283 &   8601 &  23.51 &  0.06 &  0.04 &  27.51 &  3.18 &   34 &   182 &  -11.91 \\
      &                      &                   &           &           &        &     &       &        &        &       &       &        &       &      &       &         \\
      &  UGC 8508            &  133044.4  545436 &  111.1410 &   17.9057 &  14.06 &  10 &   186 &   8601 &  23.13 &  0.04 &  0.03 &  27.15 &  2.69 &  -63 &   896 &  -13.09 \\
      &                      &                   &           &           &        &     &       &        &        &       &       &        &       &      &       &         \\
14+08 &  UGC 9240, DDO 190   &  142443.4 +443133 &   82.0084 &   26.8518 &  13.05 &  10 &   263 &   8601 &  23.21 &  0.04 &  0.02 &  27.24 &  2.80 &   -6 &   463 &  -14.19 \\
      &  UGC 8760, DDO 183   &  135050.6 +380109 &   77.7919 &   20.4675 &  14.35 &  10 &   257 &  10210 &  23.54 &  0.06 &  0.03 &  27.55 &  3.24 &  -12 &   213 &  -13.20 \\
      &  UGC 8651, DDO 181   &  133953.8 +404421 &   89.7337 &   18.5795 &  14.46 &  10 &   272 &  10210 &  23.36 &  0.05 &  0.01 &  27.40 &  3.02 &    3 &   184 &  -12.94 \\
      &  UGC 8833            &  135448.7 +355015 &   69.7137 &   21.0897 &  15.09 &  10 &   285 &  10210 &  23.50 &  0.07 &  0.02 &  27.53 &  3.20 &   16 &   258 &  -12.43 \\
      &                      &                   &           &           &        &     &       &        &        &       &       &        &       &      &       &         \\
17+06 &  NGC 784             &  020117.0  285015 &  140.9029 &   -6.3056 &  11.98 &   8 &   386 &  10210 &  24.64 &  0.05 &  0.11 &  28.58 &  5.19 &    8 &   114 &  -16.59 \\
      &  UGC 1281            &  014931.5  323517 &  136.8680 &   -2.5928 &  12.67 &   8 &   367 &  10210 &  24.59 &  0.03 &  0.09 &  28.55 &  5.13 &  -11 &   310 &  -15.88 \\
      &  KK 16               &  015520.3  275714 &  139.7495 &   -5.4137 &  16.   &  10 &   400 &  10210 &  24.78 &  0.05 &  0.14 &  28.69 &  5.47 &   22 &   330 &  -12.7  \\
      &  KK 17               &  020010.2  284953 &  140.6333 &   -6.0816 &  17.   &  10 &   360 &  10210 &  24.51 &  0.07 &  0.11 &  28.45 &  4.91 &  -18 &   282 &  -11.4  \\
      &                      &                   &           &           &        &     &       &        &        &       &       &        &       &      &       &         \\
      &  UGC 685             &  010722.4  164102 &  128.4313 &    1.6065 &  14.51 &  10 &   349 &  10210 &  24.42 &  0.03 &  0.11 &  28.36 &  4.70 &  -29 &  1257 &  -13.84 \\
      &                      &                   &           &           &        &     &       &        &        &       &       &        &       &      &       &         \\
14-14 &  NGC 1313            &  031815.4 -662951 &  283.3598 &  -28.2223 &   9.19 &   6 &   270 &  10210 &  24.25 &  0.03 &  0.21 &  28.09 &  4.15 &      &       &  -18.90 \\
      &  ESO 115-021         &  023748.1 -612018 &  282.7904 &  -25.9622 &  13.23 &   6 &   337 &  10210 &  24.49 &  0.03 &  0.05 &  28.49 &  4.99 &      &       &  -15.26 \\
      &  KK 27               &  032102.4 -661909 &  282.9130 &  -28.5528 &  16.17 &  -3 &       &  10210 &  24.19 &  0.07 &  0.15 &  28.09 &  4.15 &      &       &  -11.92 \\
      &                      &                   &           &           &        &     &       &        &        &       &       &        &       &      &       &         \\
14+14 &  ESO 154-023         &  025651.2 -543423 &  271.8077 &  -30.2306 &  12.62 &   7 &   412 &  10210 &  24.79 &  0.03 &  0.03 &  28.80 &  5.76 &  -13 &   373 &  -16.18 \\
      &  IC 1959             &  033309.0 -502442 &  261.2838 &  -36.8068 &  13.21 &   8 &   464 &  10210 &  24.88 &  0.05 &  0.02 &  28.91 &  6.06 &   39 &   443 &  -15.70 \\
      &  NGC 1311            &  032006.7 -521113 &  265.2951 &  -34.2690 &  13.09 &   9 &   398 &  10210 &  24.67 &  0.03 &  0.04 &  28.68 &  5.45 &  -27 &   311 &  -15.46 \\
      &                      &                   &           &           &        &     &       &        &        &       &       &        &       &      &       &         \\
14+19 &  UGC 3974, DDO 47    &  074152.0 +164754 &  203.0986 &  -55.5005 &  13.51 &  10 &   160 &  10210 &  25.54 &  0.04 &  0.06 &  29.53 &  8.04 &  -19 &   245 &  -16.02 \\
      &  UGC 4115            &  075702.4 +142312 &  207.0078 &  -56.2273 &  14.21 &  10 &   210 &  10210 &  25.44 &  0.05 &  0.05 &  29.44 &  7.72 &   31 &   443 &  -15.23 \\
      &  KK 65, CG 87-33     &  074232.0 +163339 &  203.3957 &  -55.6679 &  15.42 &  10 &   168 &  10210 &  25.53 &  0.07 &  0.06 &  29.52 &  8.01 &  -11 &   199 &  -14.10 \\
      &                      &                   &           &           &        &     &       &        &        &       &       &        &       &      &       &         \\
      &  UGC 3755            &  071351.6 +103119 &  206.0134 &  -63.3576 &  14.71 &  10 &   190 &  10210 &  25.47 &  0.06 &  0.17 &  29.35 &  7.41 &   11 &  1067 &  -14.64 \\
      &                      &                   &           &           &        &     &       &        &        &       &       &        &       &      &       &         \\
Dregs &  UGC 9128, DDO 187   &  141556.5 +230319 &   25.5731 &   24.3518 &  14.27 &  10 &   172 &  10210 &  22.75 &  0.05 &  0.05 &  26.75 &  2.24 &   11 &   301 &  -12.48 \\
      &  U8091, D155, GR8    &  125840.4 +141303 &  310.7381 &    4.6684 &  14.57 &  10 &   136 &   5915 &  22.64 &  0.05 &  0.05 &  26.64 &  2.13 &  -25 &   490 &  -12.07 \\
      &  KKH 86              &  135433.6 +041435 &  339.0439 &   15.4650 &  16.68 &  10 &   209 &   8601 &  23.07 &  0.13 &  0.05 &  27.07 &  2.60 &   48 &   673 &  -10.39 \\
      &  KK 230              &  140710.5 +350337 &   63.7099 &   23.5498 &  17.84 &  10 &   126 &   9771 &  22.63 &  0.08 &  0.03 &  26.65 &  2.14 &  -35 &   697 &   -8.81 \\
      &                      &                   &           &           &        &     &       &        &        &       &       &        &       &      &       &         \\
Isol  &  KKR 25              &  161347.9 +542216 &   83.8790 &   40.3703 &  17.   &  -5 &       &   8601 &  22.62 &  0.15 &  0.02 &  26.65 &  2.14 &      &       &   -9.7  \\
\enddata
\end{deluxetable}

\clearpage
\pagestyle{myheadings}
\clearpage

\begin{deluxetable}{llccccccccccc}
\rotate
\tabletypesize{\scriptsize}
\tablecaption{Properties of Groups of Dwarf Galaxies. \label{tbl-2}}
\tablewidth{0pt}
\tablehead{
\colhead{Group} & \colhead{Principal} & \colhead{No.} & \colhead{Dist.} &
\colhead{$R_I^{3D}$} &
\colhead{$V_r$}  & \colhead{$L_B$} & \colhead{$M_{\rm pm}$} &
\colhead{$M_{\rm vir}$}     & \colhead{$M_{\rm pm}/L$}  &
\colhead{$M_{\rm vir}/L$}   & \colhead{$M/L_B^{old}$}           &
\colhead{$t_{\rm x} H_0$} \\
 & \colhead{Galaxy} & & \colhead{(Mpc)} & \colhead{(Mpc)} &
\colhead{(km s$^{-1}$)} & \colhead{($10^8 L_{\odot}$)} &
\colhead{($10^{11} M_{\odot}$)} & \colhead{($10^{11} M_{\odot}$)} &
\colhead{($M_{\odot}/L_{\odot}$)} & \colhead{($M_{\odot}/L_{\odot}$)} &
\colhead{($M_{\odot}/L_{\odot}$)} &
}
\startdata
14+12 & NGC 3109 & 5 &1.37& 0.35 & 18 & 3.7 & 1.9 & 1.4 &  ~~~510 &~~360 & 1220/~300 & 0.86 \\
14+13 & NGC~~~~55& 5 &2.07& 0.28 & 36 &40.7 & 3.8 &~6.5 & ~~~~90 &\
~~160 & ~~13/~~17 & 0.33 \\
      &          & 6\tablenotemark{c} & & 0.45 & 36 &43.2 & 6.7 &~8.4 & ~~~160 &\
~~190 &      & 0.54 \\
14~+7 & NGC 4214 & 6 &2.94& 0.32 & 42 &13.7 & 5.5 & 8.:\tablenotemark{a} & ~~~400 &~~~590\tablenotemark{a} &     & 0.30 \\
14~+8 & UGC 8760 & 4 &3.06& 0.30 & 11 & 1.4 & 0.4 &~0.5 & ~~~300 &~~380 & ~250/~945 & 1.2~ \\
17~+6 & NGC~~~784& 4 &5.2~& 0.26 & 17 &10.5 & 1.3 &~1.7 & ~~~120 &~~160 & ~330/1110 & 0.68 \\
14+14 & ESO154-23& 3 &5.8~& 0.38 & 35 &10.0 & 7.3 &10.4 & ~~~730 &~1040 &      & 0.48 \\
14+19 & UGC 3974 & 3 &7.9~& 0.31 & 26 & 6.6 & 4.0 &~2.:\tablenotemark{a} & ~~~600 &~~300\tablenotemark{a} &      & 0.52 \\
      &          & 4\tablenotemark{b} &    & 0.67 & 22 & 7.7 & 3.5 &~2.:\tablenotemark{a} & ~~~450 &~~300\tablenotemark{a} & 1060/2040 & 1.3~ \\
\tableline
Dregs & DDO~155  & 4 &2.28& 0.57 & 37 & 0.3 & 13. &16.  &44000 &56000 &     & 0.66 \\
\enddata


\tablenotetext{a}{Virial mass estimate biased low: close pair}
\tablenotetext{b}{Including UGC 3755 in group}
\tablenotetext{c}{Including IC 5152 in group}


\end{deluxetable}

\begin{figure}
\figurenum{1}
\caption{
The 20 newly observed galaxies.
Top: Antlia and UGC 9128; 2nd row: UGC 8651 and UGC 8833;
3rd row: UGC 8760 and KK 27; bottom: NGC 1313 and UGC 685.
\label{fig:montage}}
\end{figure}

\begin{figure}
\figurenum{1}
\caption{
continued.
Top pair: ESO 115-021 and UGC 1281; middle pair: NGC 784 and NGC 1311;
bottom pair: KK 17 and KK 16.}
\end{figure}

\begin{figure}
\figurenum{1}
\caption{
continued
Top pair: ESO 154-023 and IC 1959; middle pair: UGC 3755 and UGC 3974;
bottom pair: UGC 4115 and KK 65.}
\end{figure}

\begin{figure}
\figurenum{2}
\plotone{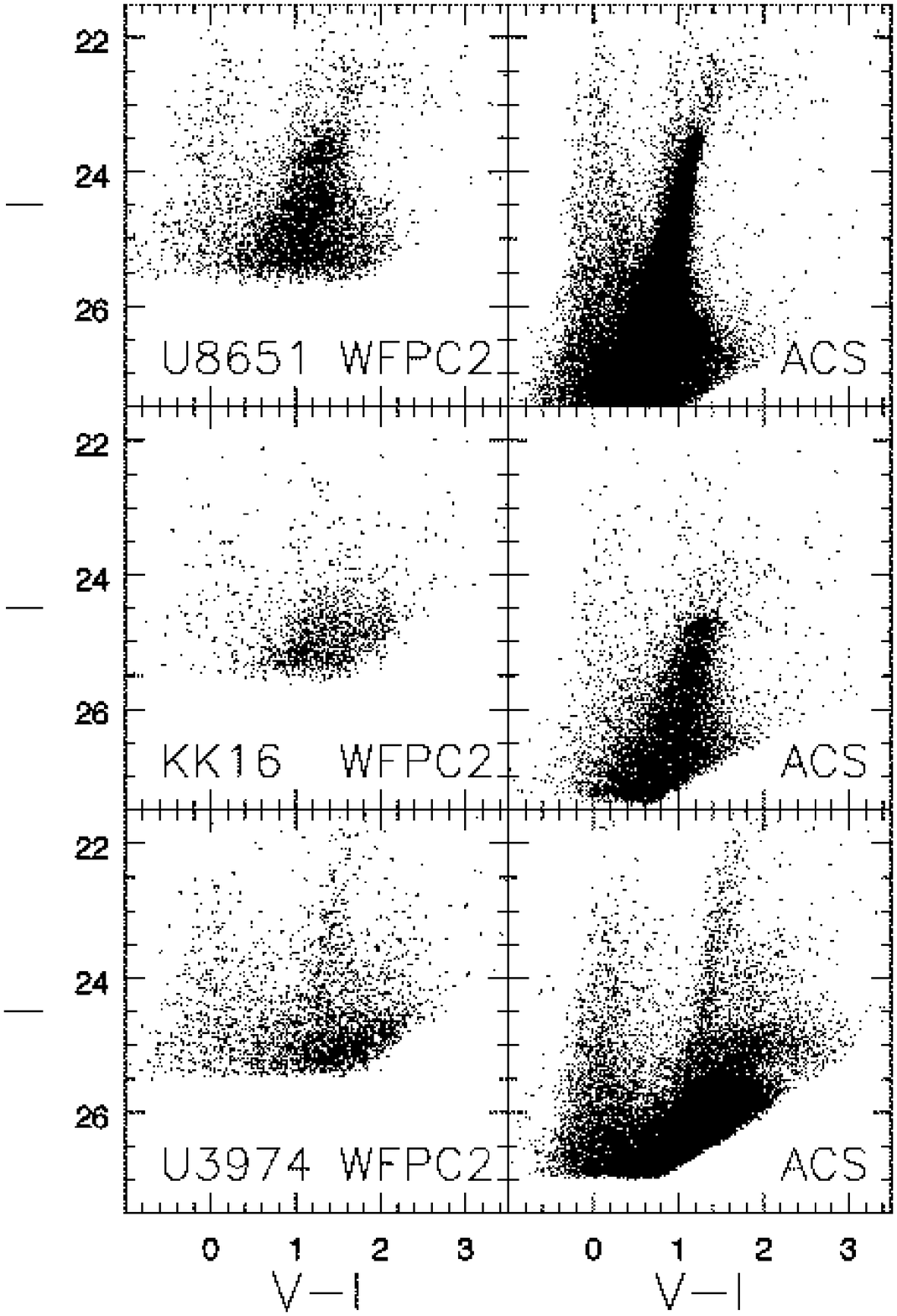}
\caption{
Color--magnitude diagrams for three dwarf galaxies. 
{\it Top:} UGC~8651,
TRGB at $I=23.36$, distance 3.02 Mpc. {\it Middle:} KK~16,  TRGB at $I=24.78$,
distance 5.5 Mpc. 
{\it Bottom:} UGC~3974, TRGB at $I=25.54$, distance 8.0~Mpc.
The data in the left panels were obtained with 600s exposures in the F814W
filter with WFPC2 during SNAP program 8601.  The data in the right panels
were obtained with, respectively, 1209s, 1226s, and 1226s exposures in the 
same filter with ACS 
during GO program 10210. In each case, the data were obtained in a single 
HST orbit.
The advantage of ACS over WFPC2 is apparent.
\label{fig:wfpc2-acs}}
\end{figure}

\begin{figure}
\figurenum{3}
\plotone{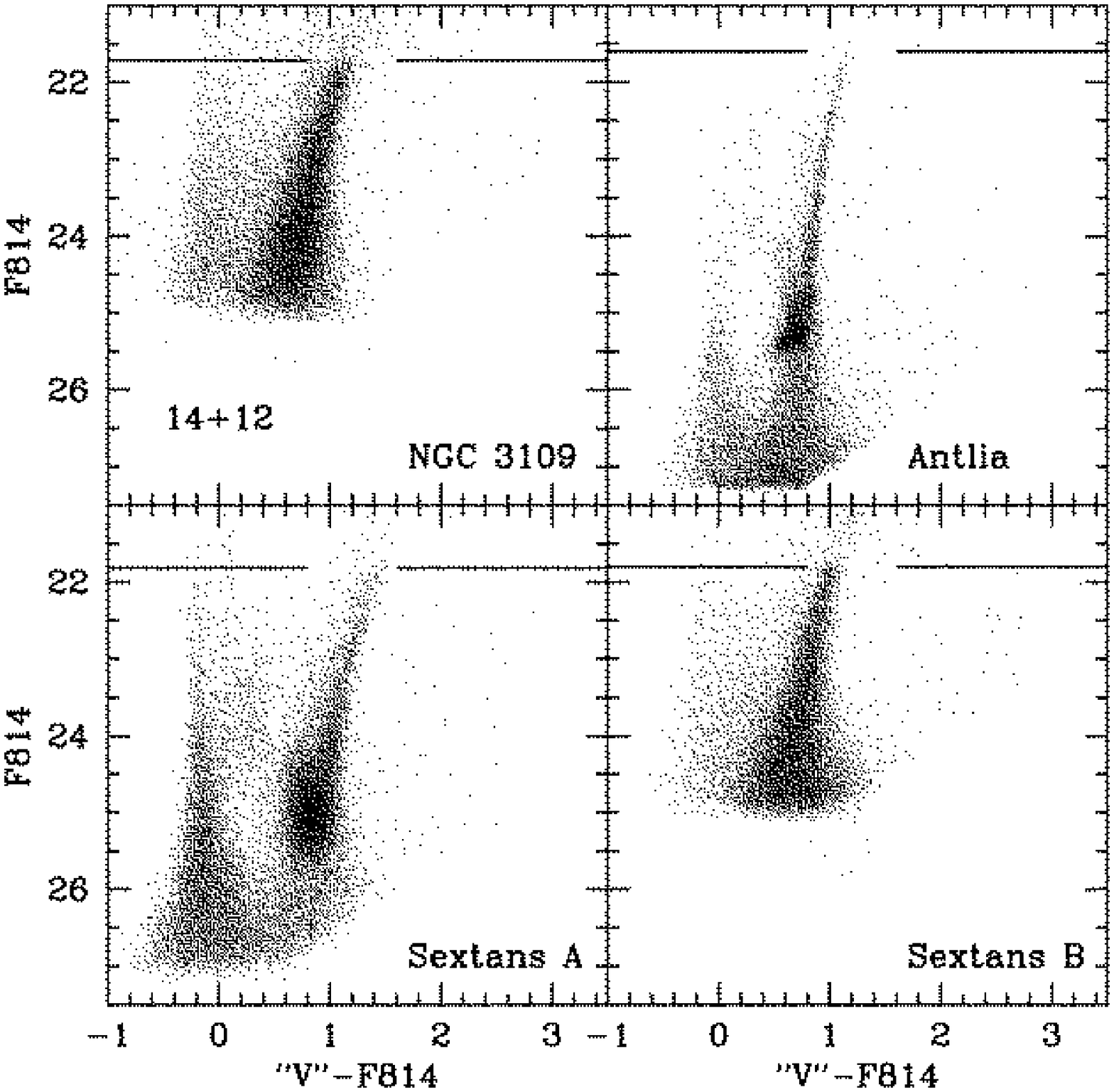}
\caption{
14+12 Association. The NGC~3109 and Sex~B data come from HST program 8601 with
600s exposures in the F814W filter.  The Sex~A data comes from
a 9,600s exposure in the F814W filter in program 7496.  The Antlia data
was acquired in 1174s at F814W with ACS in our program 10210.  In the case
of Sextans~A, the ``V'' image is obtained with the F555W filter.  In the
other cases, ``V'' is obtained with the F606W filter.
The data
for Sex~A and Antlia is superior but the TRGB is well defined in all 
cases (marked by horizontal lines in each panel). 
\label{fig:14+12}}
\end{figure}

\clearpage

\begin{figure}
\figurenum{4}
\plotone{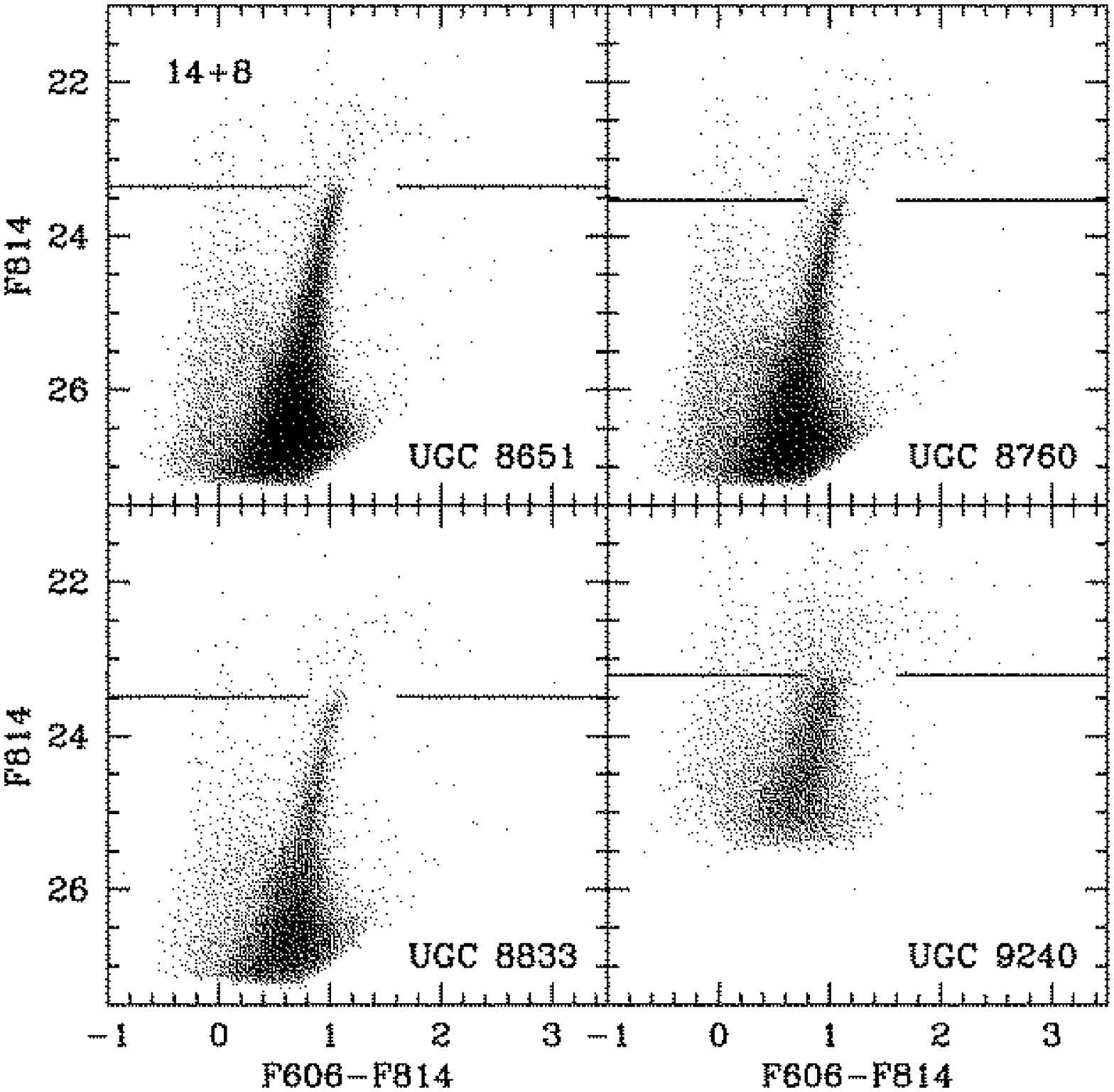}
\caption{
14+8 Association. The data for UGC~8651, UGC~8760, and UGC~8833 come from our
program 10210 with F814W exposures of 1209s, 1209s, and 1189s, respectively.
The data for UGC~9240 is provided by program 8601 with a 600s exposure
in the F814W band.  Horizontal lines mark the TRGB.
\label{fig:14+8}}
\end{figure}

\clearpage

\begin{figure}
\figurenum{5}
\centerline{\psfig{figure=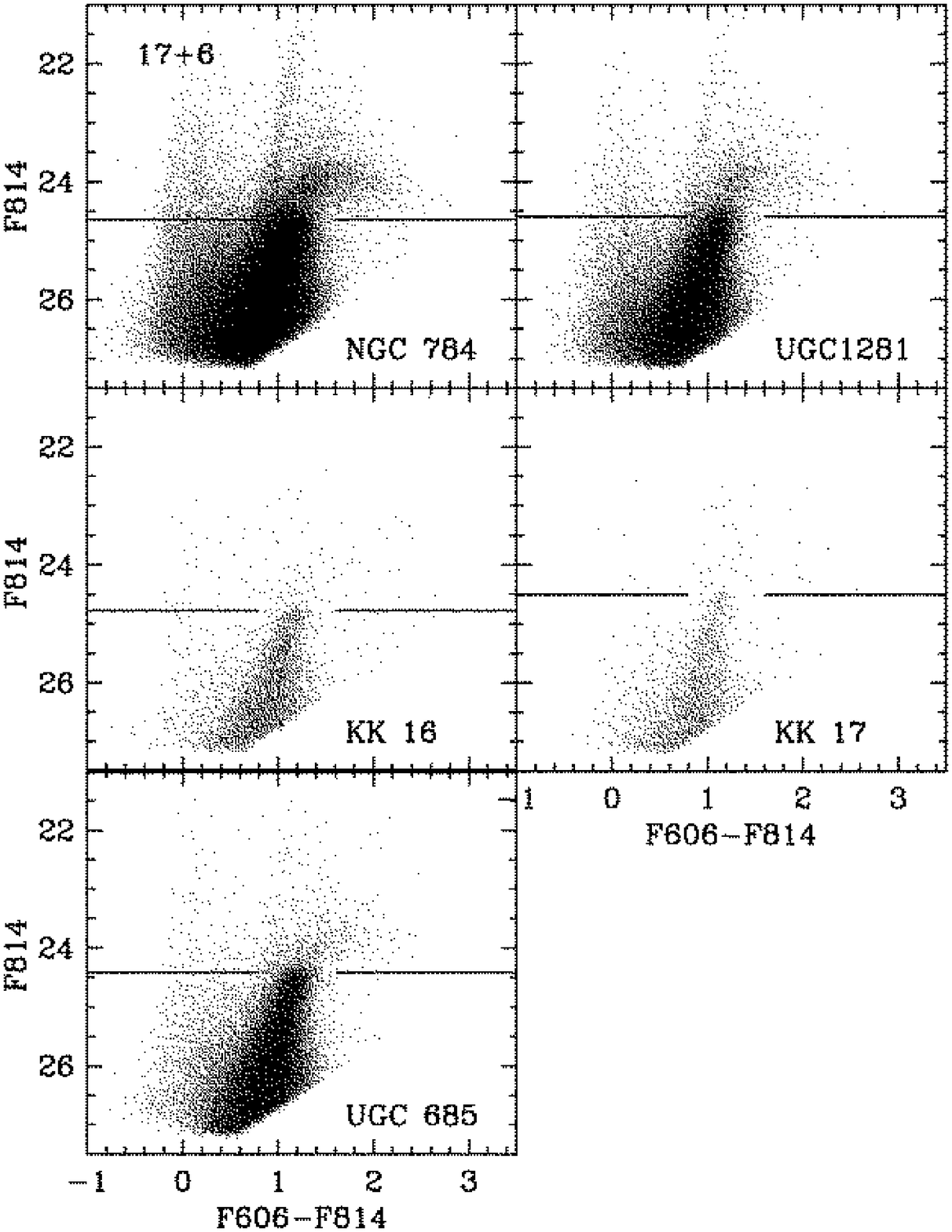,width=0.9\textwidth,angle=0}}
\caption{
17+6 Association. All data from our program 10210 with F814W exposures of 1226s.
TRGB marked by horizontal lines.
\label{fig:17+6}}
\end{figure}

\begin{figure}
\figurenum{6}
\plotone{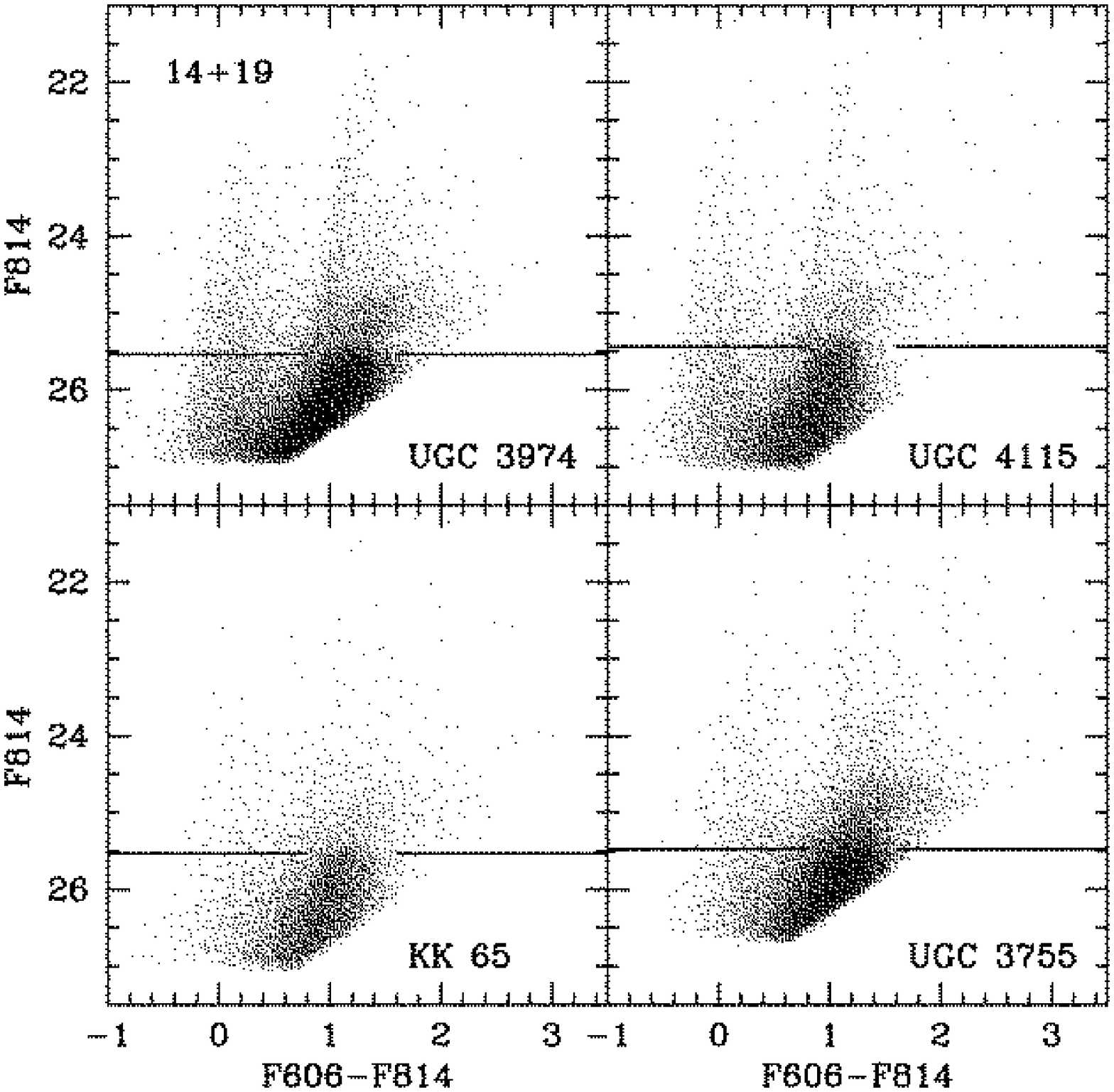}
\caption{
14+19 Association.  All data from our program 10210 with F814W exposures of 1226s.
TRGB marked by horizontal lines.
\label{fig:14+19}}
\end{figure}

\begin{figure}
\figurenum{7}
\centerline{\psfig{figure=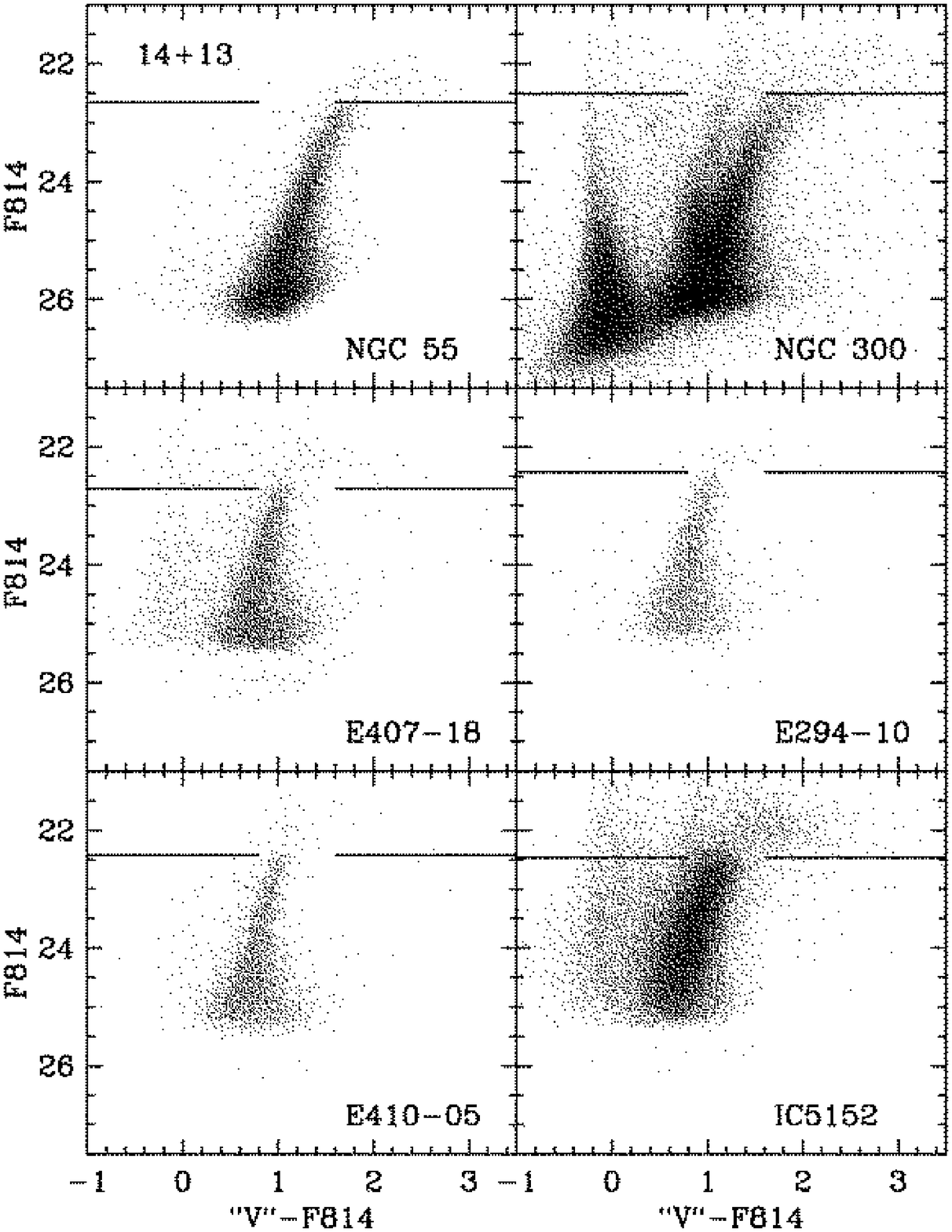,width=0.9\textwidth,angle=0}}
\caption{
14+13 Association.  All data acquired with WFPC2.  NGC~55: 2,500s in F814W; 
NGC~300: 1,000s in F814W; ESO~407-18, ESO~294-10, ESO~410-05, and IC~5152: 
600s in F814W.
For NGC~55, the ``V'' filter is F555W. For the others, 
the ``V'' filter is F606W.
TRGB marked by horizontal lines.
\label{fig:14+13}}
\end{figure}

\begin{figure}
\figurenum{8}
\centerline{\psfig{figure=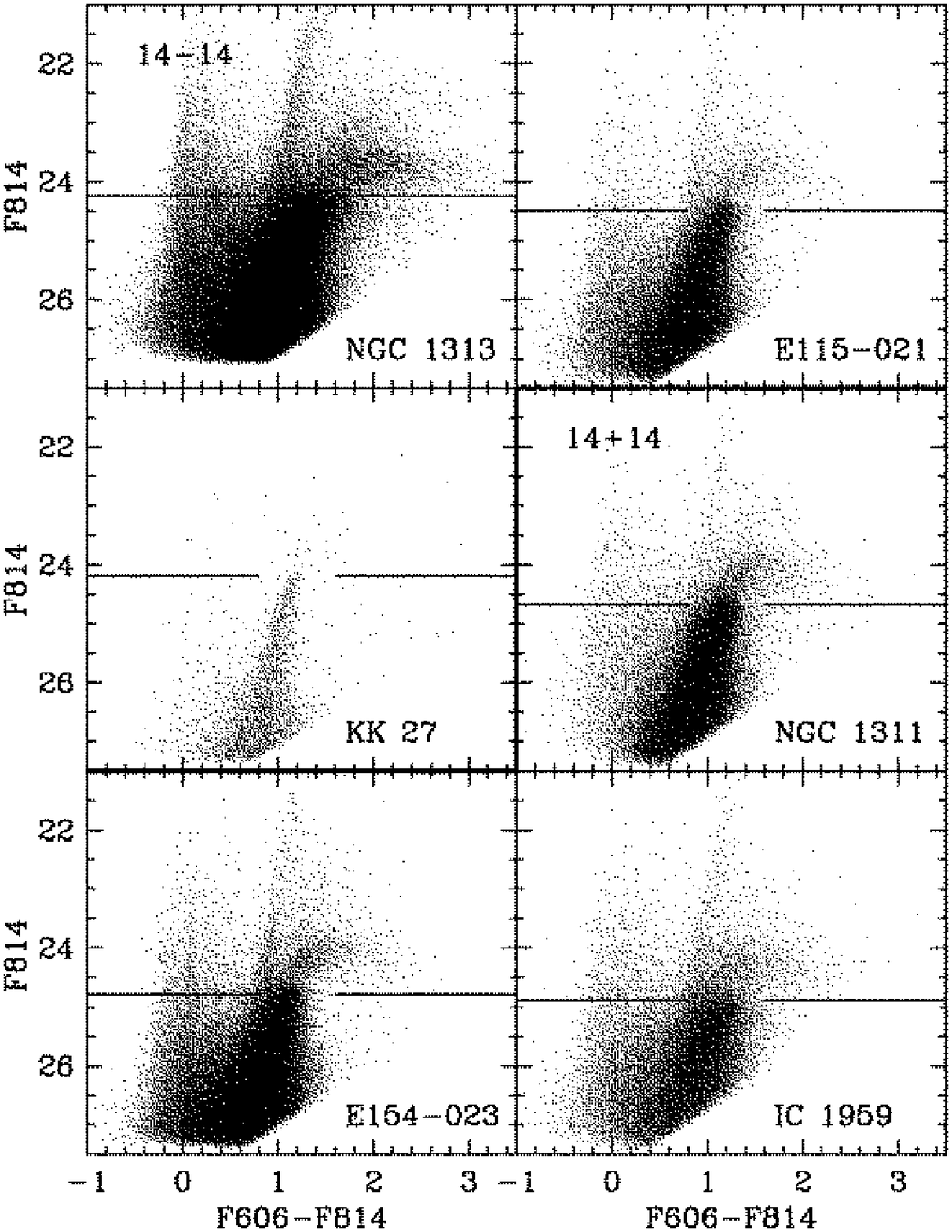,width=0.9\textwidth,angle=0}}
\caption{
14-14 Group (above heavy outline) and 14+14 Association (below heavy 
boundary).  All data from our program 10210.
TRGB marked by horizontal lines.
\label{fig:14+14}}
\end{figure}

\begin{figure}
\figurenum{9}
\centerline{\psfig{figure=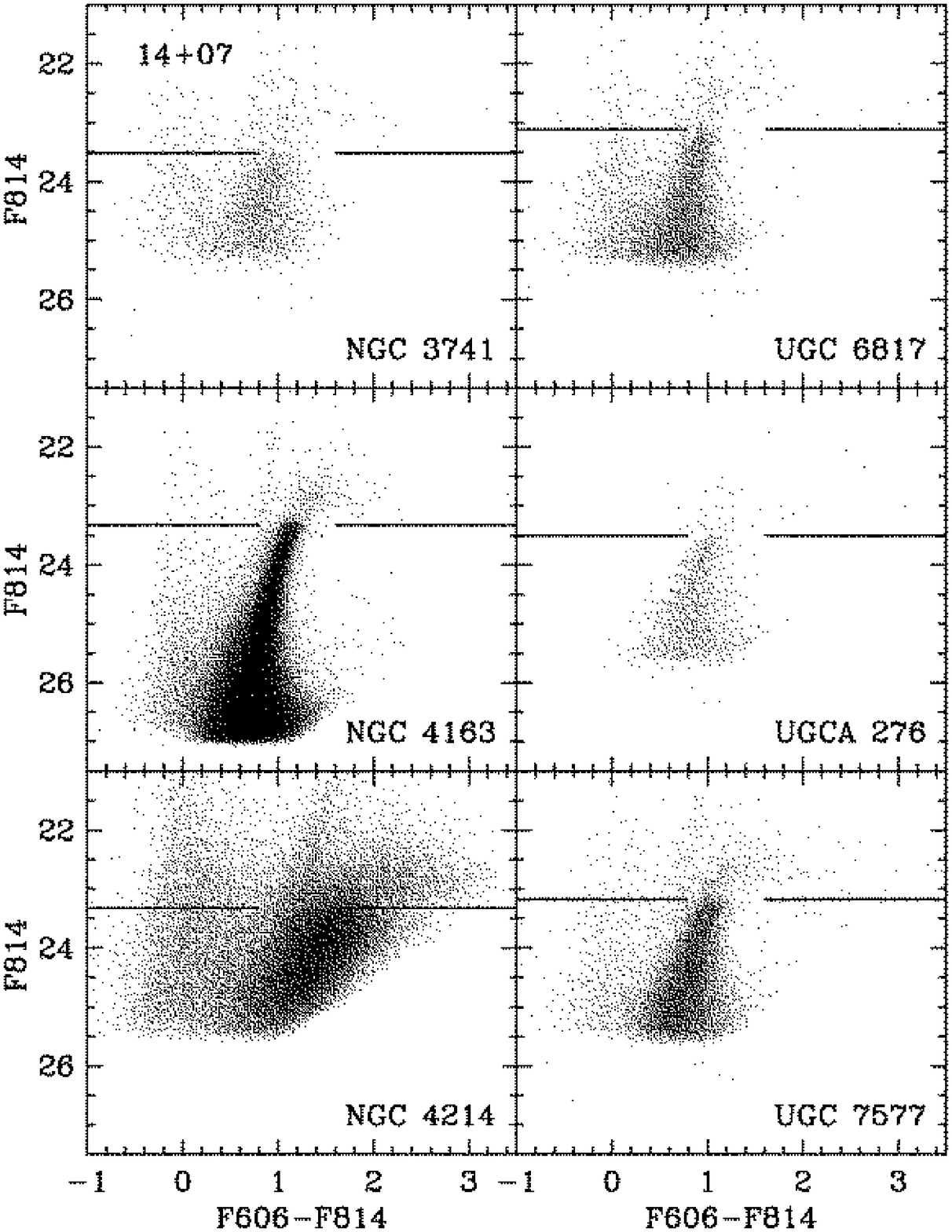,width=0.9\textwidth,angle=0}}
\caption{
14+7 Association members.  
All these CMDs are derived from HST archival material.
UGC~8508 is an outlying candidate member.  
KKR~25 is the galaxy most notable for its isolation.  
\label{fig:14+7}}
\end{figure}

\begin{figure}
\figurenum{10}
\centerline{\psfig{figure=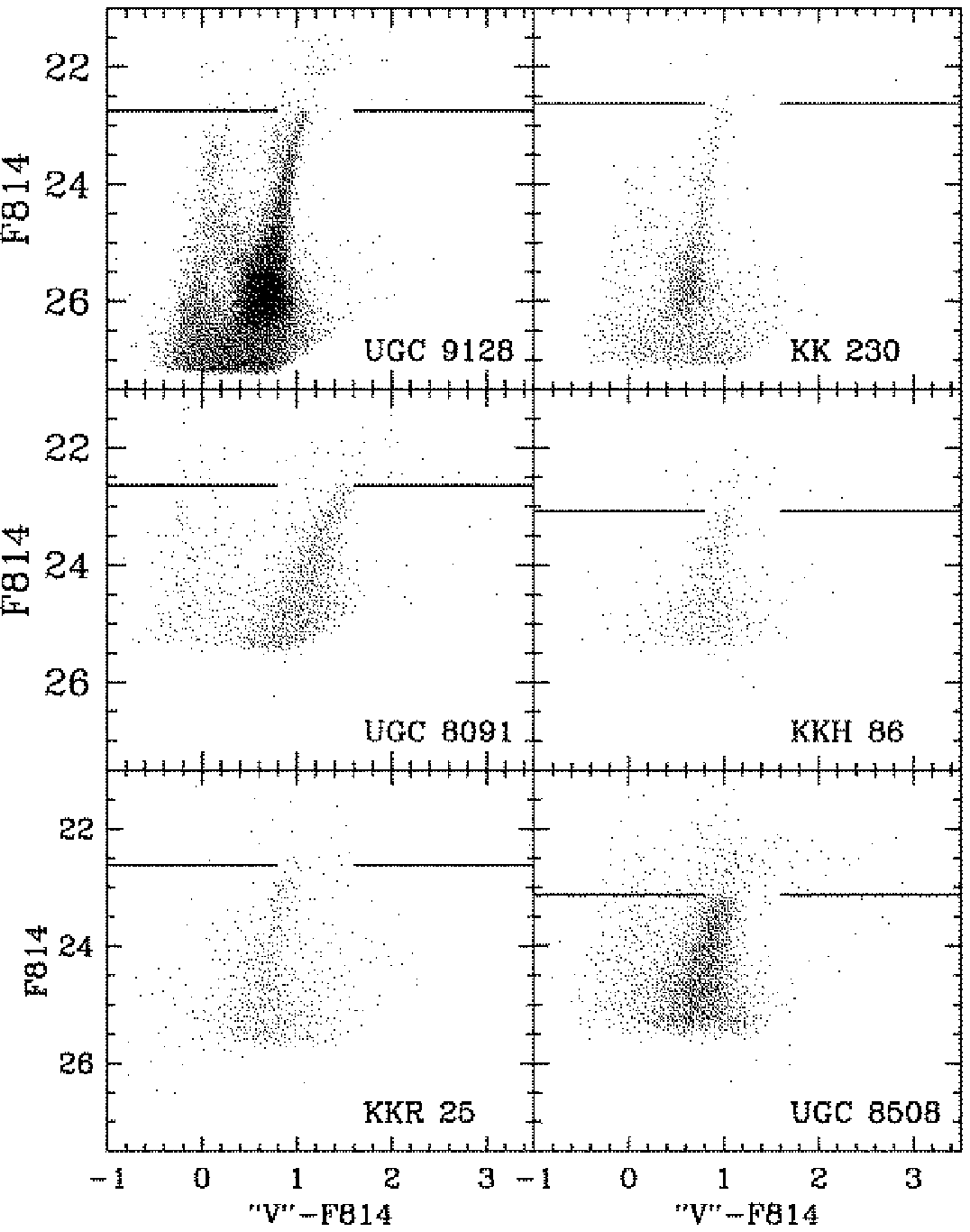,width=0.9\textwidth,angle=0}}
\caption{
Dregs in vicinity of DDO~155.  CMD for UGC~9128 is from our ACS program 10210.
The other data are from the HST archive.  In the case of UGC~8091 = DDO~155
the $V$ data is obtained in the F555W filter; otherwise F606W.
\label{fig:dregs}}
\end{figure}

\begin{figure}
\figurenum{11}
\plotone{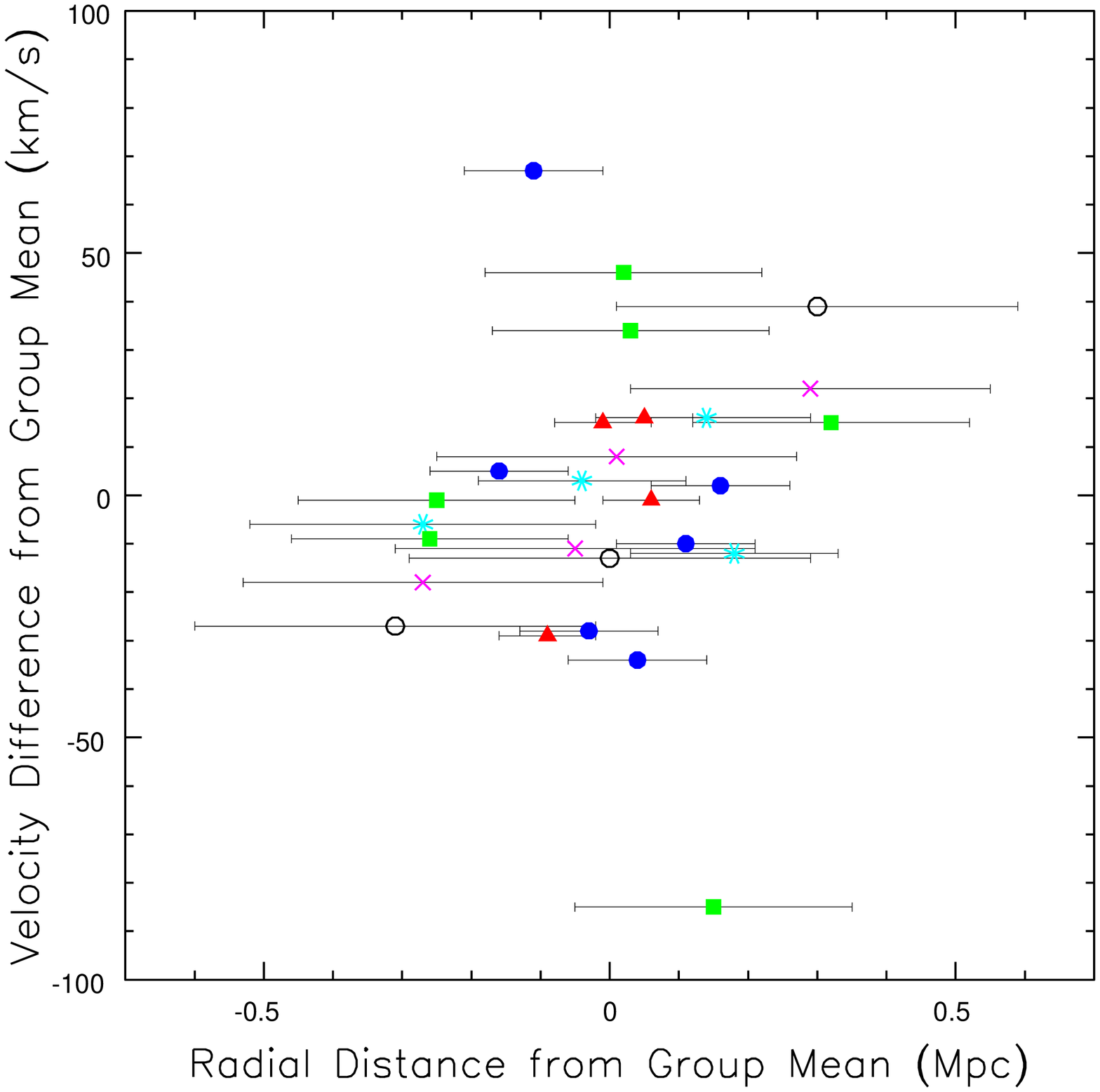}
\caption{
Velocity difference compared with association mean as a function of 
distance compared with the association mean.  Triangles: 14+12 Group;
eight-point stars: 14 +8 Group; filled circles: 14+13 Group; crosses: 17+6
Group; squares: 14+7 Group; open circles: 14+14 Group.
\label{fig:dv6}}
\end{figure}

\begin{figure}
\figurenum{12}
\plotone{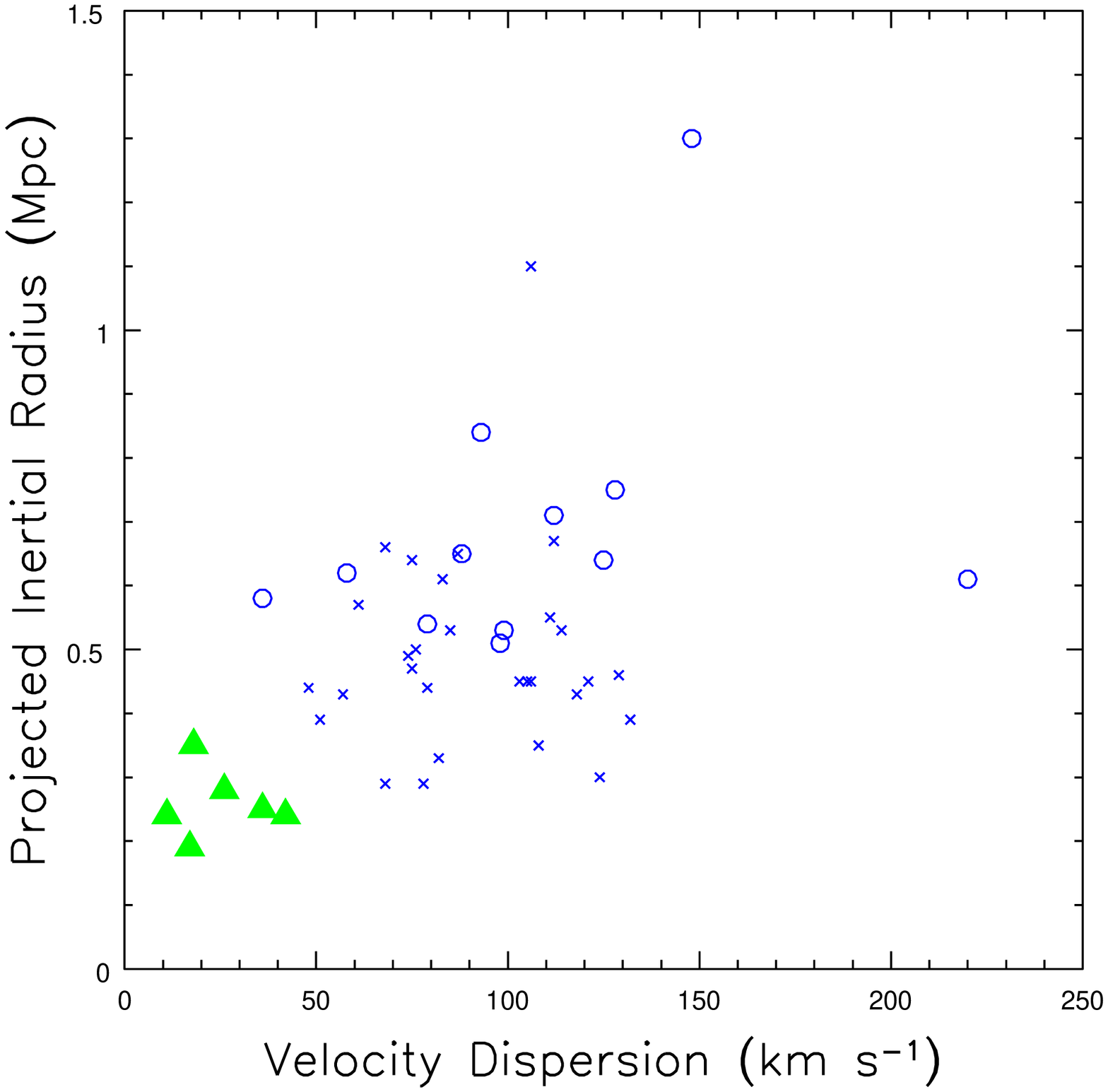}
\caption{
Projected inertial radius vs velocity dispersion for associations of dwarfs
and spiral dominated groups.  Triangles: associations of dwarfs.  Circles:
spiral dominated groups with at least 6 galaxies with $M_B < -17$. Crosses:
less populated groups.
\label{fig:rv}}
\end{figure}

\begin{figure}
\figurenum{13}
\plotone{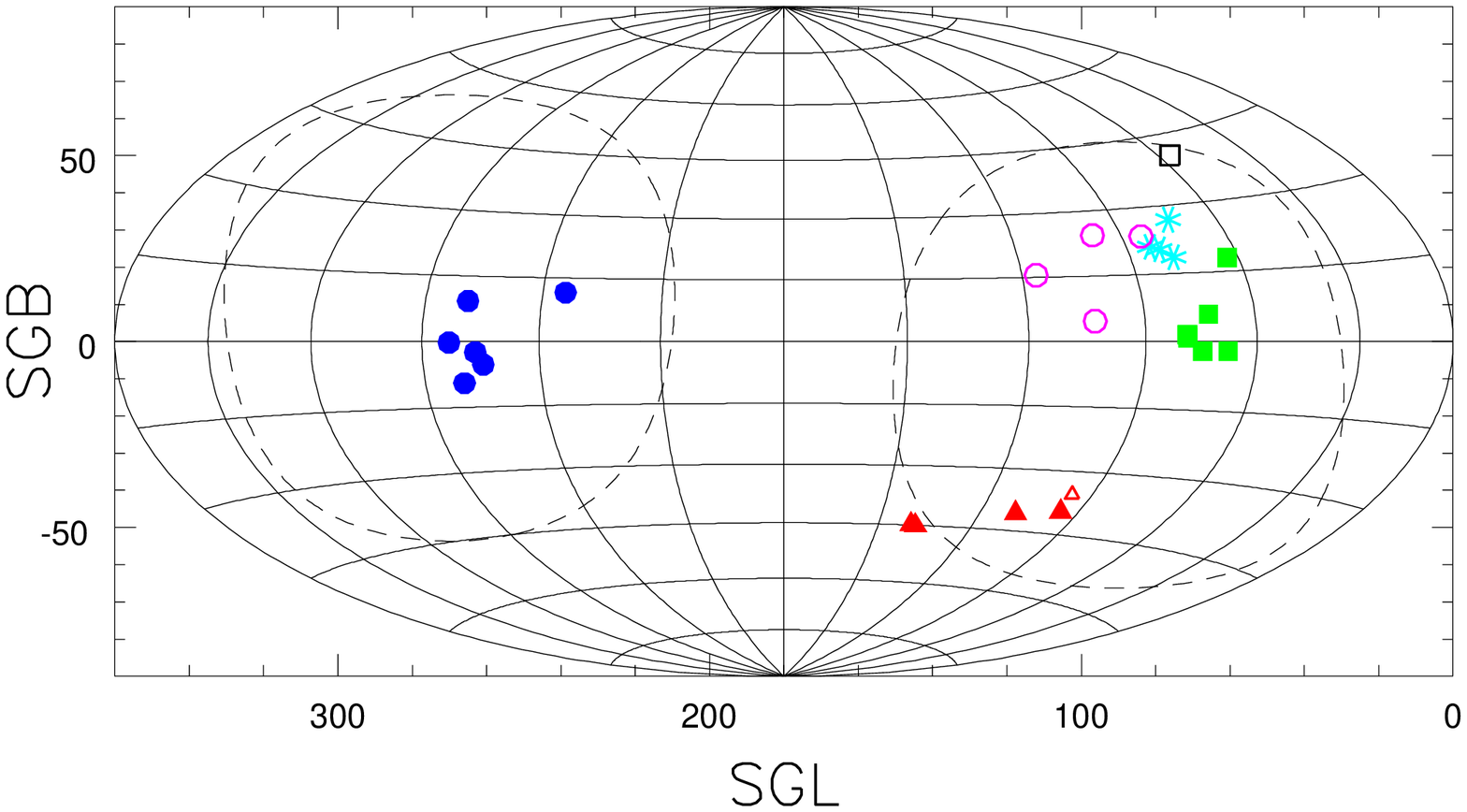}
\caption{
Distribution in supergalactic coordinates of galaxies with accurately
known distances between 1.1~Mpc and 3.2~Mpc and at high Galactic latitude.
Triangles: 14+12 Group; filled circles: 14+13 Group; filled squares: 14+7
Group; eight-point stars: 14+8 Group; open circles: dregs; open square:
KKR~25.  Dashed contours: $b = \pm 30$.  The small open triangle locates
KKH~60 which is suspected to lie within this volume.  Other candidates lie
in the region around the 14+7 Association.
\label{fig:shell}}
\end{figure}

\begin{figure}
\figurenum{14}
\centerline{\psfig{figure=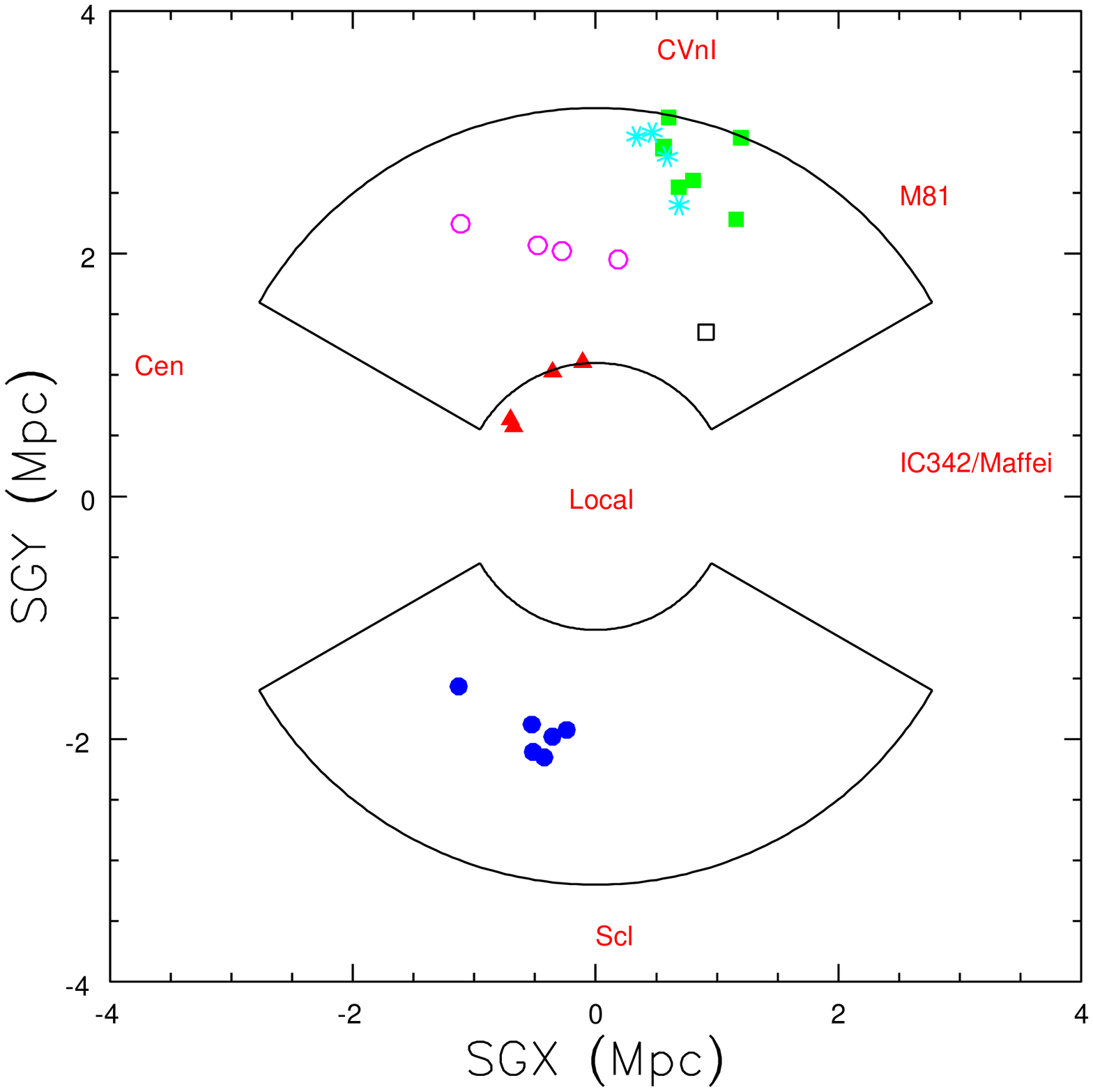,width=0.6\textwidth,angle=0}}
\centerline{\psfig{figure=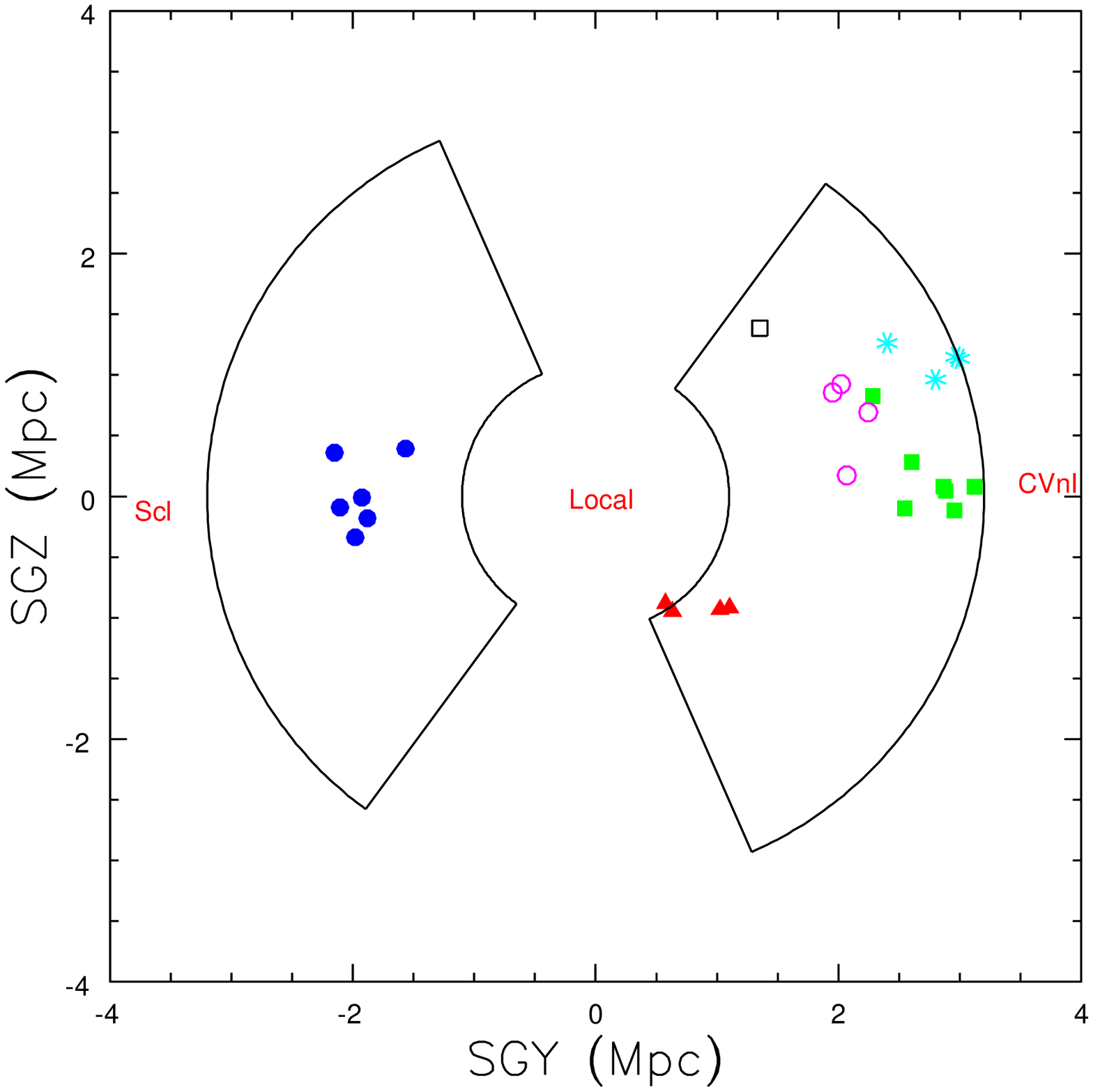,width=0.6\textwidth,angle=0}}
\caption{
Locations of 26 galaxies with accurate distances in the shell between
1.1 and 3.2~Mpc and $\vert b \vert > 30$.  The borders are shown for the
central planes of each plot; ie, at SGZ=0 and SGX=0 respectively.  
Symbols as in Fig.~\ref{fig:shell}.
Locations of nearest groups beyond 3.2~Mpc are indicated by labels.
\label{fig:xyz}}
\end{figure}

\begin{figure}
\figurenum{15}
\plotone{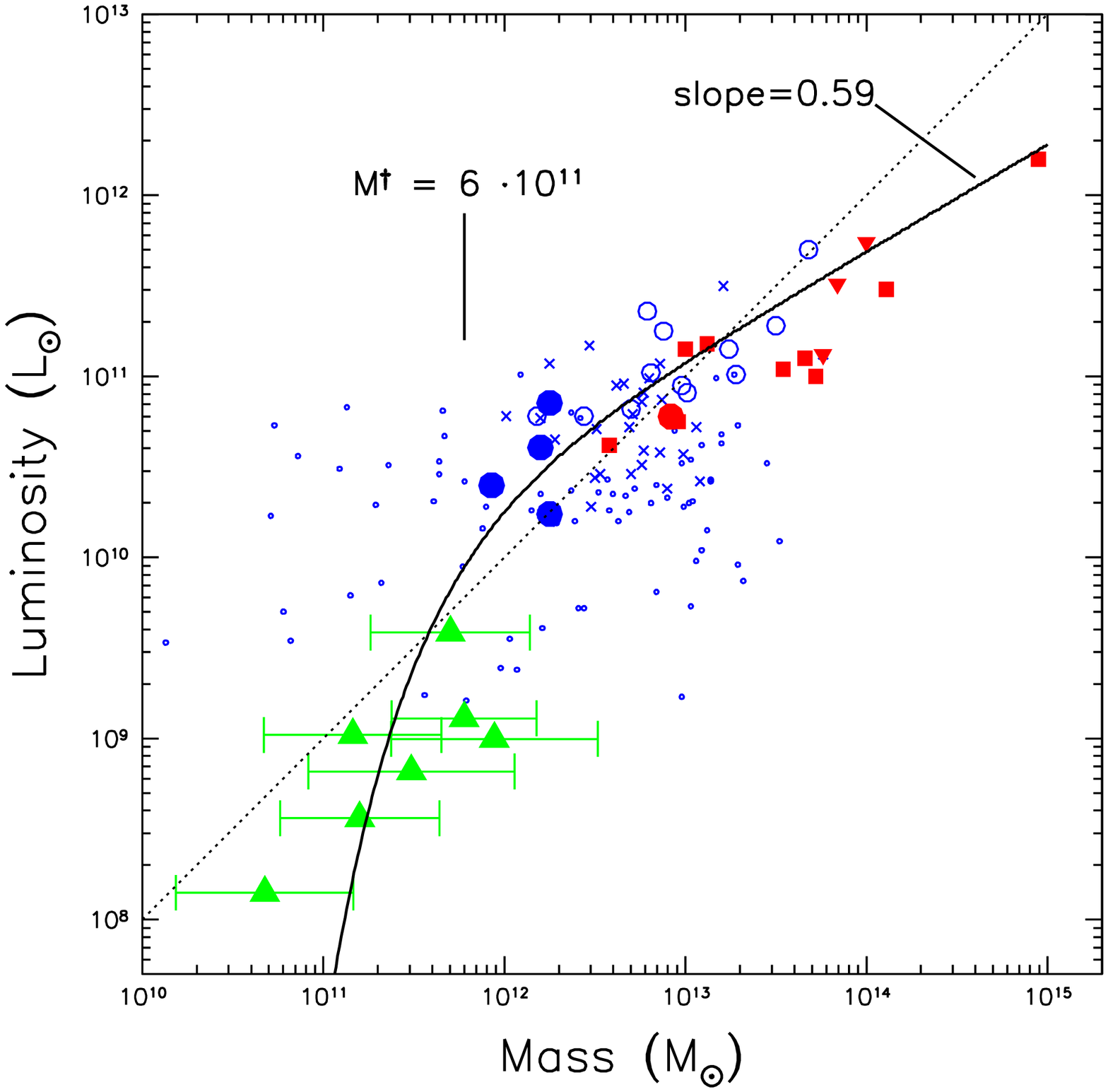}
\caption{
Luminosity as a function of mass for groups and associations in the Local
Supercluster.  Squares and inverted triangles: groups dominated by E/S0
galaxies.  Open circles: groups dominated by spirals (large filled circles: 
Local, M81, CVnI, and Cen~A groups).  Crosses and tiny circles: small groups 
(down to pairs).
Large triangles with error bars: associations of dwarf galaxies discussed
in this paper.  Dotted line: $M/L_B=100$.  Solid curve: fit to data.  This
figure is carried over from T05 with updated information on the associations
of dwarfs.  See that paper for details.
\label{fig:ml}}
\end{figure}

\begin{figure}
\figurenum{16}
\plotone{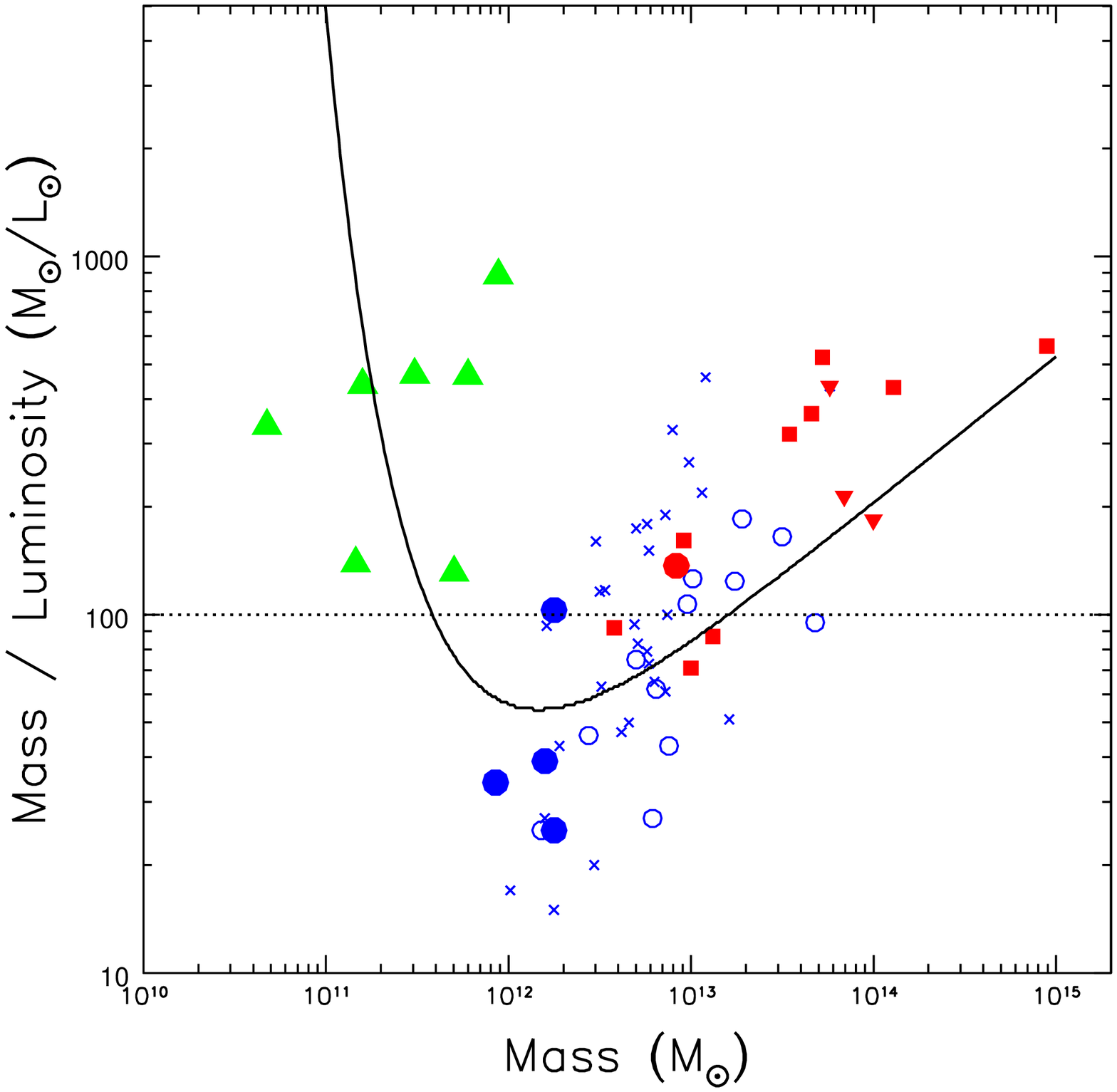}
\caption{
$M/L_B$ as a function of mass for groups and associations.  Symbols and lines
have the same meanings as in the previous plot.  The previous tiny circles 
representing groups of 2--4 members have been excluded.
\label{fig:mlm}}
\end{figure}

\end{document}